\documentclass[12pt]{article}
\usepackage[margin=1in]{geometry}
\usepackage{graphicx}
\usepackage{amsmath, amssymb}
\usepackage[numbers,sort&compress]{natbib}
\usepackage{booktabs}
\usepackage[colorinlistoftodos]{todonotes}
\setuptodonotes{size=\scriptsize}
\usepackage[printonlyused]{acronym}
\usepackage[hidelinks]{hyperref}
\usepackage{pdflscape}
\usepackage{tikz}
\usetikzlibrary{shapes.geometric, arrows.meta, positioning, fit, backgrounds, calc, shadows.blur}
\usepackage{xcolor}
\definecolor{msgreen}{RGB}{0,164,0}
\definecolor{msblue}{RGB}{0,120,212}

\usepackage{listings}
\definecolor{codegreen}{rgb}{0.25,0.5,0.35}
\definecolor{codegray}{rgb}{0.5,0.5,0.5}
\definecolor{codepurple}{rgb}{0.58,0,0.82}
\definecolor{codeblue}{rgb}{0.0,0.4,0.7}
\definecolor{codebg}{rgb}{0.97,0.97,0.97}
\lstdefinestyle{pythonstyle}{
    backgroundcolor=\color{codebg},
    commentstyle=\color{codegreen}\itshape,
    keywordstyle=\color{codeblue}\bfseries,
    numberstyle=\tiny\color{codegray},
    stringstyle=\color{codepurple},
    basicstyle=\ttfamily\scriptsize,
    breakatwhitespace=false,
    breaklines=true,
    captionpos=b,
    keepspaces=true,
    numbers=left,
    numbersep=5pt,
    showspaces=false,
    showstringspaces=false,
    showtabs=false,
    tabsize=4,
    frame=single,
    framerule=0.5pt,
    rulecolor=\color{codegray},
    xleftmargin=2.5em,
    framexleftmargin=2em,
    aboveskip=1.5em,
    belowskip=1em,
    language=Python,
    morekeywords={self, True, False, None, as, with, yield, async, await},
}
\usepackage{tcolorbox}
\tcbuselibrary{listings, breakable, skins}
\newtcblisting{codeblock}[2][]{
    listing only,
    breakable,
    enhanced,
    colback=codebg,
    colframe=codegray,
    left=2em,
    top=0.5em,
    bottom=0.5em,
    listing options={style=pythonstyle},
    title={#2},
    #1
}

\definecolor{usercolor}{HTML}{5D6D7E}       
\definecolor{factorycolor}{HTML}{8B6B7A}    
\definecolor{instancecolor}{HTML}{6B7D68}   
\definecolor{registrycolor}{HTML}{A69580}   
\definecolor{msgcolor}{HTML}{4A5B6A}        

\definecolor{inputbg}{HTML}{F5F0E8}
\definecolor{inputborder}{HTML}{A69580}
\definecolor{inputtext}{HTML}{6B5B4F}
\definecolor{classicalbg}{HTML}{E8ECF0}
\definecolor{classicalborder}{HTML}{7A8B99}
\definecolor{classicaltext}{HTML}{4A5B6A}
\definecolor{classicalbox}{HTML}{5D6D7E}
\definecolor{classicalboxborder}{HTML}{4A5B6A}
\definecolor{quantumcircuitbg}{HTML}{E8EDE8}
\definecolor{quantumcircuitborder}{HTML}{7D9079}
\definecolor{quantumcircuittext}{HTML}{5A6B58}
\definecolor{quantumcircuitbox}{HTML}{6B7D68}
\definecolor{quantumcircuitboxborder}{HTML}{5A6B58}
\definecolor{simbg}{HTML}{F0E8EC}
\definecolor{simborder}{HTML}{A08090}
\definecolor{simtext}{HTML}{6B5060}
\definecolor{simbox}{HTML}{8B6B7A}
\definecolor{simboxborder}{HTML}{6B5060}
\definecolor{devtoolsub}{HTML}{9A7A8A}
\definecolor{devtoolsubborder}{HTML}{8B6B7A}
\definecolor{postprocessbg}{HTML}{F5EDE8}
\definecolor{postprocessborder}{HTML}{A08070}
\definecolor{postprocesstext}{HTML}{6B5545}
\definecolor{postprocessbox}{HTML}{B8866B}
\definecolor{postprocessboxborder}{HTML}{8B6650}
\hypersetup{
    pdfauthor={QDK/Chemistry Authors},
    pdftitle={QDK/Chemistry: A Modular Toolkit for Quantum Chemistry Applications}
}
\usepackage{authblk}

\usepackage{url}

\newcommand{\qdkchem}{\mbox{QDK/Chemistry}}

\newlength{\innerboxmargin}
\title{QDK/Chemistry: A Modular Toolkit for Quantum Chemistry Applications\thanks{Authors are listed in alphabetical order.}}

\author[1]{Nathan~A.~Baker}
\author[1]{Brian~Bilodeau}
\author[1]{Chi~Chen}
\author[1]{Yingrong~Chen}
\author[2]{Marco~Eckhoff}
\author[1]{Alexandra~Efimovskaya}
\author[1]{Piero~Gasparotto}
\author[2]{Puck~van~Gerwen}
\author[1]{Rushi~Gong}
\author[1]{Kevin~Hoang}
\author[1]{Zahra~Hooshmand}
\author[1]{Andrew~J.~Jenkins}
\author[1]{Conrad~S.~N.~Johnston}
\author[1]{Run~R.~Li}
\author[1]{Jiashu~Liang}
\author[1]{Hongbin~Liu}
\author[1]{Alexis~Mills}
\author[2]{Maximilian~M\"orchen}
\author[1]{George~Nishibuchi}
\author[1]{Chong~Sun}
\author[1]{Bill~Ticehurst}
\author[1]{Matthias~Troyer}
\author[2]{Jan~P.~Unsleber}
\author[1]{Stefan~Wernli}
\author[1]{David~B.~Williams-Young\thanks{Corresponding author: davidwillia@microsoft.com}}
\author[1]{Boqin~Zhang}
\affil[1]{Microsoft Quantum, Redmond, WA, USA}
\affil[2]{Microsoft Quantum, Copenhagen, Denmark}

\date{\today}

\begin{document}

\maketitle

\begin{abstract}
We present \qdkchem{}, a software toolkit for quantum chemistry workflows targeting quantum computers.
The toolkit addresses a key challenge in the field: while quantum algorithms for chemistry have matured considerably, the infrastructure connecting classical electronic structure calculations to quantum circuit execution remains fragmented.
\qdkchem{} provides this infrastructure through a modular architecture that separates data representations from computational methods, enabling researchers to compose workflows from interchangeable components.
In addition to providing native implementations of targeted algorithms in the quantum-classical pipeline, the toolkit builds upon and integrates with widely used open-source quantum chemistry packages and quantum computing frameworks through a plugin system, allowing users to combine methods from different sources without modifying workflow logic.
This paper describes the design philosophy, current capabilities, and role of \qdkchem{} as a foundation for reproducible quantum chemistry experiments.
\end{abstract}

\newpage
\tableofcontents

\section{Introduction}
\label{sec:introduction}

Quantum chemistry is among the most compelling application domains for quantum computers~\cite{Lanyon2010,McArdle2020,Dalzell2025,Hoefler2023}.
The promise of efficiently simulating strongly correlated electronic systems, where classical methods face exponential scaling, has motivated extensive algorithmic development over the past two decades~\cite{Reiher2017,Cao2019,Lee2021}.
Algorithms such as \ac{QPE}~\cite{Abrams1999,Aspuru-Guzik2005,Svore2014} and \ac{VQE}~\cite{Peruzzo2014,McClean2016} provide frameworks for extracting ground state energies and properties, while advances in error correction bring fault-tolerant implementations within reach~\cite{Aasen2025,Alexeev2025}.

However, a quantum algorithm alone does not solve a chemistry problem.
The path from a molecular structure to a meaningful energy estimate involves a substantial classical preprocessing pipeline: selecting a basis set, computing molecular integrals, performing mean-field calculations, choosing an active space that captures the essential correlation physics while conforming to hardware constraints, and mapping classical quantities (e.g., wavefunctions and Hamiltonians) onto qubit operators.
These classical stages prepare inputs for quantum execution which, much like classical computation, require extensive compilation and optimization to translate high-level algorithms into hardware-executable circuits.
After quantum execution, classical postprocessing must aggregate measurement statistics into observable estimates.
To connect with practical chemistry problems, multiple observable estimates need to be aggregated.
Furthermore, in order to be comparable, all of the included calculations must have been computed in a consistent manner.

Consider, for example, a reaction energy profile.
Computing this profile to obtain the reaction energy and barrier requires energy evaluations that include at least the starting, transition, and final states.
Therefore, the solution of chemistry problems is not a single calculation, but rather a campaign executing multiple workflows, with each starting from structures with unique geometries and then proceeding through the aforementioned stages.
Each of these stages involves methodological choices for the scientist, where the most optimal choice for each stage depends on the specific molecule, hardware platform, and scientific question at hand.

In practice, researchers often assemble these stages using \textit{ad hoc} scripts that bridge multiple software packages.
This approach often encounters several problems:
\begin{itemize}
    \item Data formats differ between packages, requiring custom conversion code that is error-prone and rarely reusable.
    \item Methodological variations are difficult to track, hindering reproducibility.
    \item Comparing alternative approaches---i.e., different active space selectors, qubit mappings, or state preparation strategies---requires rewriting pipeline logic rather than simply swapping components.
    \item As quantum algorithms and hardware evolve rapidly, inflexible pipelines quickly become obsolete.
\end{itemize}
These challenges are not new~\cite{DiFelice2023, Lehtola2023, Blum2024} and not unique to quantum chemistry or quantum applications; similar issues arise across computational science domains where complex workflows span multiple software systems.
The classical computational chemistry community has responded with modular frameworks for workflow management and interoperability, including workflow engines~\cite{Pizzi2016, Larsen2017, Huber2020, Uhrin2021, Huber2021}, standardized data schemas and archives~\cite{Smith2021, Scheidgen2023}, plugin-based software architectures~\cite{Kowalski2021, Richard2023, Barnes2024}, and integrated platforms~\cite{Alvarez2015, Weymuth2024, Berquist2024}.
In parallel, the quantum computing community has developed frameworks specifically addressing the chemistry-to-circuit pipeline: OpenFermion~\cite{McClean2020} provides foundational data structures and fermion-to-qubit transformations; Tequila~\cite{Kottmann2021} offers high-level abstractions for variational algorithms with backend agnosticism; Tangelo~\cite{Senicourt2022} emphasizes end-to-end workflows with problem decomposition; and full-stack frameworks like Qiskit~\cite{Javadi-Abhari2024}, PennyLane~\cite{Bergholm2022}, and Cirq~\cite{Cirq2024} provide circuit-level infrastructure with chemistry-specific extensions.
These efforts have significantly advanced the field, yet each addresses a particular slice of the workflow: some focus on operator transformations, others on variational optimization, and still others on circuit compilation or hardware abstraction.

Despite a preponderance of successes, modular frameworks face a recurring tension: as they mature to address broader use cases, their abstractions often grow in complexity until significant expertise is required to understand, extend, or maintain them.
The interfaces designed to simplify component substitution can become barriers to entry when they accumulate layers of indirection.
For quantum chemistry workflows targeting quantum computers, this tension is particularly acute because the field evolves rapidly---both the quantum algorithms and the classical methods that support them---and extensibility must remain practical rather than merely architectural.

With these considerations in mind, we have developed \qdkchem{}\footnote{\url{https://github.com/microsoft/qdk-chemistry}}: a permissively licensed, open-source, modular toolkit for the development and exploration of quantum chemistry workflows targeting quantum computers focusing on robustness, practical extensibility, and ease-of-use.
The toolkit combines native implementations with capabilities from widely adopted open-source packages in the quantum chemistry and quantum computing communities, enabling researchers to leverage mature, community-tested software alongside purpose-built components.
Rather than implementing a fixed pipeline, the toolkit defines interfaces that decouple workflow stages, enabling researchers to substitute methods at any stage without modifying surrounding code.
This design serves the diverse needs of users across the quantum chemistry community:
\begin{itemize}
    \item Application scientists can systematically explore methodological alternatives to determine which approaches work best for their molecules and scientific questions of interest.
    \item Method developers, ranging from classical electronic structure theorists to quantum algorithm developers, can integrate their implementations into a mature software stack without reconstructing the surrounding infrastructure.
    \item Educators and students gain access to a coherent framework for understanding the complete workflow from molecular specification to quantum measurement.
\end{itemize}
By serving the needs of these different user groups, \qdkchem{} aims to accelerate progress toward practical quantum utility in the field of electronic structure calculations.
Appendix~\ref{appendix:example-code} provides a complete code example demonstrating the workflow from molecular specification through quantum phase estimation.

\qdkchem{} is an application library layer within the broader Microsoft \ac{QDK}~\cite{QDK}, which provides the underlying infrastructure for quantum algorithm development and execution.
The \ac{QDK} encompasses the Q\#, Qiskit, Cirq, and QASM programming languages for expressing quantum algorithms, compilers that translate high-level programs into optimized circuits, high-performance simulators for algorithm validation and debugging, and integrated development tooling within Visual Studio Code \cite{VSCode}.
\qdkchem{} builds on this foundation, leveraging \ac{QDK}'s compilation and simulation infrastructure while providing the domain-specific data structures, algorithms, and workflows needed for quantum chemistry applications.
Through its plugin architecture, the toolkit also integrates with the broader quantum software ecosystem, enabling researchers to combine \ac{QDK} capabilities with external tools as their workflows require.
This layered design allows chemistry researchers to focus on scientific questions while benefiting from ongoing advances in both the underlying \ac{QDK} stack and the wider community of quantum chemistry and quantum computing software.
The remainder of this paper discusses the design and software infrastructure underlying \qdkchem{}, describes its current capabilities, and outlines its role in the broader ecosystem of quantum chemistry and quantum applications software.

\section{Software Design}
\label{sec:design}

The design of \qdkchem{} reflects several principles that emerged from experience building quantum chemistry workflows for quantum computers.
These principles guide architectural decisions throughout the codebase.

\subsection{Separation of data and algorithms}

A quantum chemistry workflow produces a sequence of intermediate results: a molecular geometry becomes a set of orbitals, which become a Hamiltonian, which becomes a qubit operator, and so on.
In many codebases, these quantities are intertwined with the methods that produce them, making it difficult to substitute one method for another or to inspect intermediate results.
\qdkchem{} enforces a strict separation:
\begin{itemize}
    \item Data classes represent intermediate quantities as immutable, self-contained objects.
    \item Algorithm classes are stateless transformations that consume data objects and produce new ones.
\end{itemize}
This separation yields a simple dataflow model: each stage takes well-defined inputs and produces well-defined outputs.
Changing the method at one stage does not affect the behaviors at other stages (see Listing~\ref{lst:workflow-example} for a concrete illustration).

Immutability of data classes, while seemingly restrictive, simplifies provenance: the output of a calculation depends only on its inputs and configuration, not on hidden state accumulated from previous operations.
Algorithm configuration (e.g., method-specific options such as convergence thresholds, iteration limits) follows a similar pattern.
Each algorithm exposes a type-safe settings object with easily discoverable parameters that locks after execution, preventing accidental modification.
Both data and settings objects support serialization, enabling checkpointing and ensuring that calculations can be reproduced from archived inputs, and allowing for integrations with database and workflow management systems.

\subsection{Uniform interfaces for algorithm substitution}

Many workflow stages admit multiple implementations.
\Ac{SCF} calculations can use different convergence algorithms or different software backends; orbital localization can use different cost functions or proceed by different optimization strategies; qubit mapping can use different encodings with various trade-offs targeting different hardware; \textit{et cetera}.
In an ever-evolving landscape of methods, algorithms, hardware, and software, users should be able to switch implementations without rewriting pipeline code.
This flexibility matters because different users have different goals: some are primarily interested in developing new algorithms, while others want to benchmark existing methods against their own problems of choice.

\qdkchem{} achieves this through factory-based instantiation for algorithm interfaces.
Each algorithm type has an associated factory that maintains a registry of available implementations.
Conceptually, an algorithm interface is simply a name for all algorithms that require the same input data and produce the same output data.
Users request an implementation by name; the factory returns an object conforming to the common interface.
Client code operates on interfaces rather than concrete types, so substitution is transparent.

\begin{figure}[t]
\centering
\begin{tikzpicture}
    \def\outerwidth{\textwidth}
    \def\outerheight{9cm}
    
    \draw[thick] (0,0) rectangle (\outerwidth, \outerheight);
    
    
    \def\groupleft{\innerboxmargin}
    \def\groupbottom{5.2cm}  
    \def\groupwidth{4cm}
    \def\groupheight{3.5cm}
    \def\innerpadding{0.2cm}
    \def\smallboxheight{1.1cm}
    \def\smallboxgap{0.2cm}
    \def\labelspace{0.5cm}
    
    \draw[dashed] (\groupleft, \groupbottom) rectangle +(\groupwidth, \groupheight);
    
    \node[below] at (\groupleft + \groupwidth/2, \groupbottom + \groupheight - 0.1cm) {Client Layer};
    
    \draw[thick] (\groupleft + \innerpadding, \groupbottom + \smallboxheight + \smallboxgap + \innerpadding) 
        rectangle +(\groupwidth - 2*\innerpadding, \smallboxheight)
        node[midway] {Algorithm Query};
    
    \draw[thick] (\groupleft + \innerpadding, \groupbottom + \innerpadding) 
        rectangle +(\groupwidth - 2*\innerpadding, \smallboxheight)
        node[midway] {Algorithm Object};
    
    \def\groupbottomB{\innerboxmargin}
    
    \def\groupheightB{4.0cm}
    \draw[dashed] (\groupleft, \groupbottomB) rectangle +(\groupwidth, \groupheightB);
    
    \node[below] at (\groupleft + \groupwidth/2, \groupbottomB + \groupheightB - 0.1cm) {Data Layer};
    
    \def\databoxheight{0.8cm}
    \def\datalabelspace{0.9cm}  
    \def\databoxgap{0.25cm}
    
    \draw[thick] (\groupleft + \innerpadding, \groupbottomB + \groupheightB - \datalabelspace - \databoxheight) 
        rectangle +(\groupwidth - 2*\innerpadding, \databoxheight)
        node[midway] {Orbitals};
    
    \draw[thick] (\groupleft + \innerpadding, \groupbottomB + \groupheightB - \datalabelspace - 2*\databoxheight - \databoxgap) 
        rectangle +(\groupwidth - 2*\innerpadding, \databoxheight)
        node[midway] {QubitHamiltonian};
    
    \draw[thick] (\groupleft + \innerpadding, \groupbottomB + \groupheightB - \datalabelspace - 3*\databoxheight - 2*\databoxgap) 
        rectangle +(\groupwidth - 2*\innerpadding, \databoxheight)
        node[midway] {\ldots};
    
    \draw[thick, <->, line width=1.5pt] (\groupleft + \groupwidth/2 - 0.4cm, \groupbottom + \innerpadding) 
        -- (\groupleft + \groupwidth/2 - 0.4cm, \groupbottomB + \groupheightB);
    \draw[thick, <->, line width=1.5pt] (\groupleft + \groupwidth/2, \groupbottom + \innerpadding) 
        -- (\groupleft + \groupwidth/2, \groupbottomB + \groupheightB);
    \draw[thick, <->, line width=1.5pt] (\groupleft + \groupwidth/2 + 0.4cm, \groupbottom + \innerpadding) 
        -- (\groupleft + \groupwidth/2 + 0.4cm, \groupbottomB + \groupheightB);
    
    \def\groupwidthC{9cm}
    \def\groupCbottom{5.2cm}  
    \def\groupheightC{3.5cm}  
    
    \draw[dashed] (\outerwidth - \innerboxmargin - \groupwidthC, \groupCbottom) 
        rectangle (\outerwidth - \innerboxmargin, \groupCbottom + \groupheightC);
    
    \node[below left] at (\outerwidth - \innerboxmargin - 0.1cm, \groupCbottom + \groupheightC - 0.1cm) {Factory Layer};
    
    \def\hboxwidth{3.6cm}
    \def\hboxgap{2.0cm}
    \def\factoryboxbottom{\groupbottom + \smallboxheight/2 + \smallboxgap/2 + \innerpadding}
    \draw[thick] (\outerwidth - \innerboxmargin - \groupwidthC + \innerpadding, \factoryboxbottom) 
        rectangle +(\hboxwidth, \smallboxheight)
        node[midway] {Factory Interface};
    
    \draw[thick] (\outerwidth - \innerboxmargin - \innerpadding - \hboxwidth, \factoryboxbottom) 
        rectangle +(\hboxwidth, \smallboxheight)
        node[midway] {Registry};
    
    \draw[thick, <->, line width=1.5pt, msblue] (\outerwidth - \innerboxmargin - \groupwidthC + \innerpadding + \hboxwidth, \factoryboxbottom + \smallboxheight/2 + 0.15cm) 
        -- (\outerwidth - \innerboxmargin - \innerpadding - \hboxwidth, \factoryboxbottom + \smallboxheight/2 + 0.15cm);
    \draw[thick, <->, line width=1.5pt, msgreen] (\outerwidth - \innerboxmargin - \groupwidthC + \innerpadding + \hboxwidth, \factoryboxbottom + \smallboxheight/2 - 0.15cm) 
        -- (\outerwidth - \innerboxmargin - \innerpadding - \hboxwidth, \factoryboxbottom + \smallboxheight/2 - 0.15cm);
    
    \draw[dashed, ->, line width=1.5pt, msblue] (\groupleft + \groupwidth - \innerpadding, \groupbottom + \smallboxheight + \smallboxgap + \innerpadding + \smallboxheight/2 + 0.15cm) 
        -- (\outerwidth - \innerboxmargin - \groupwidthC, \groupbottom + \smallboxheight + \smallboxgap + \innerpadding + \smallboxheight/2 + 0.15cm);
    \draw[thick, <-, line width=1.5pt, msblue] (\groupleft + \groupwidth - \innerpadding, \groupbottom + \smallboxheight + \smallboxgap + \innerpadding + \smallboxheight/2 - 0.15cm) 
        -- (\outerwidth - \innerboxmargin - \groupwidthC, \groupbottom + \smallboxheight + \smallboxgap + \innerpadding + \smallboxheight/2 - 0.15cm)
        node[midway, below] {Metadata};
    
    \draw[dashed, ->, msgreen, line width=1.5pt] (\groupleft + \groupwidth - \innerpadding, \groupbottom + \innerpadding + \smallboxheight/2 + 0.15cm) 
        -- (\outerwidth - \innerboxmargin - \groupwidthC, \groupbottom + \innerpadding + \smallboxheight/2 + 0.15cm);
    \draw[thick, <-, msgreen, line width=1.5pt] (\groupleft + \groupwidth - \innerpadding, \groupbottom + \innerpadding + \smallboxheight/2 - 0.15cm) 
        -- (\outerwidth - \innerboxmargin - \groupwidthC, \groupbottom + \innerpadding + \smallboxheight/2 - 0.15cm)
        node[midway, below] {Instance};
    
    \def\groupwidthD{9cm}
    \def\groupheightD{4.0cm}
    \def\treeboxtopbottom{3.8cm}  
    
    \draw[dashed] (\outerwidth - \innerboxmargin - \groupwidthD, \innerboxmargin) 
        rectangle (\outerwidth - \innerboxmargin, \innerboxmargin + \groupheightD);
    
    \node[below left] at (\outerwidth - \innerboxmargin - 0.1cm, \innerboxmargin + \groupheightD - 0.1cm) {Algorithm Layer};
    
    \def\treenodewidth{2.2cm}
    \def\parentnodewidth{3.8cm}
    \def\treenodeheight{\smallboxheight}
    \def\parentbottom{2.0cm}
    \def\parentcenterx{\outerwidth - \innerboxmargin - \groupwidthD/2}
    
    \draw[thick] (\parentcenterx - \parentnodewidth/2, \innerboxmargin + \parentbottom) 
        rectangle +(\parentnodewidth, \treenodeheight)
        node[midway] {Algorithm Interface};
    
    \def\childbottom{\innerpadding}
    \def\childspacing{2.6cm}
    
    \draw[thick] (\parentcenterx - \childspacing - \treenodewidth/2, \innerboxmargin + \childbottom) 
        rectangle +(\treenodewidth, \treenodeheight)
        node[midway] {QDK};
    
    \draw[thick] (\parentcenterx - \treenodewidth/2, \innerboxmargin + \childbottom) 
        rectangle +(\treenodewidth, \treenodeheight)
        node[midway] {Qiskit};
    
    \draw[thick] (\parentcenterx + \childspacing - \treenodewidth/2, \innerboxmargin + \childbottom) 
        rectangle +(\treenodewidth, \treenodeheight)
        node[midway] {\ldots};
    
    \draw[thick] (\parentcenterx, \innerboxmargin + \parentbottom) -- (\parentcenterx - \childspacing, \innerboxmargin + \childbottom + \treenodeheight);
    \draw[thick] (\parentcenterx, \innerboxmargin + \parentbottom) -- (\parentcenterx, \innerboxmargin + \childbottom + \treenodeheight);
    \draw[thick] (\parentcenterx, \innerboxmargin + \parentbottom) -- (\parentcenterx + \childspacing, \innerboxmargin + \childbottom + \treenodeheight);
    
    \draw[dashed, msgreen, line width=1.5pt] (\outerwidth - \innerboxmargin - \innerpadding - \hboxwidth/2, \factoryboxbottom) 
        -- (\outerwidth - \innerboxmargin - \innerpadding - \hboxwidth/2, \innerboxmargin + \groupheightD + 0.3cm)
        -- (\parentcenterx, \innerboxmargin + \groupheightD + 0.3cm)
        -- (\parentcenterx, \innerboxmargin + \parentbottom + \treenodeheight);
\end{tikzpicture}
\caption{Architecture of \qdkchem{}, illustrating the separation of client code, algorithm interfaces, factory registries, and data representations.}
\label{fig:architecture}
\end{figure}
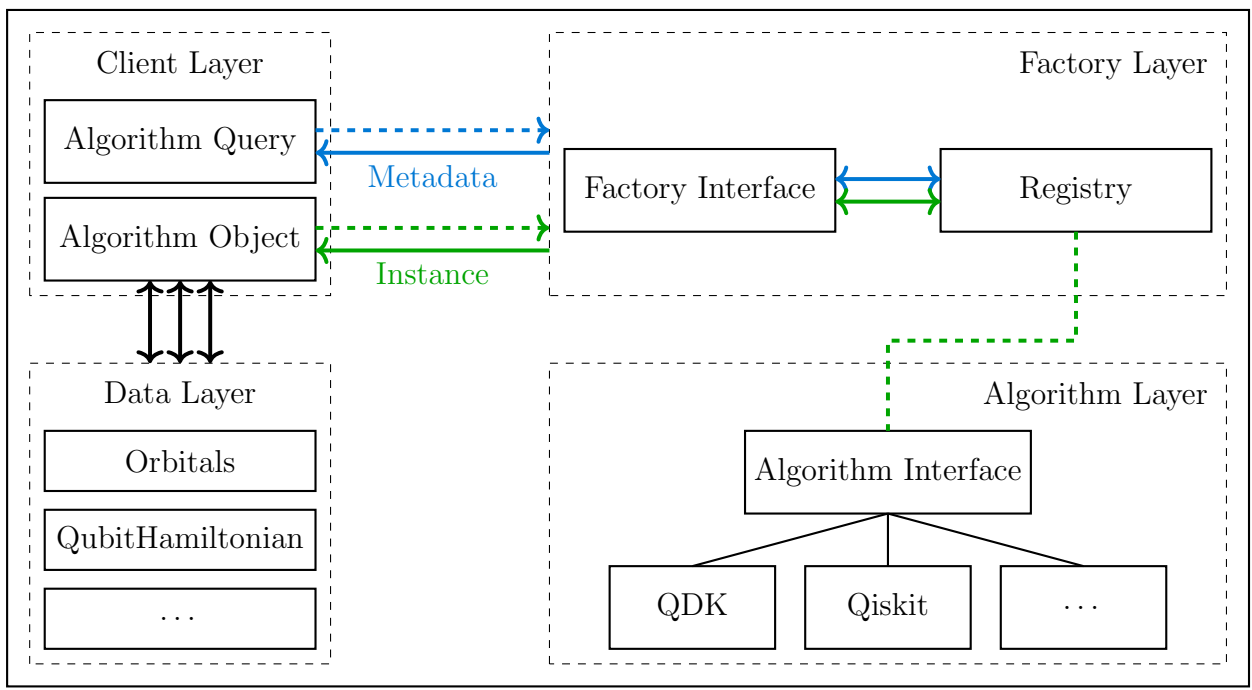

This pattern also supports discoverability.
Users can query a factory at runtime to list available implementations, facilitating the exploration of alternatives.
New implementations, whether developed internally or contributed by the community, integrate seamlessly once registered with the appropriate factory.
Figure~\ref{fig:architecture} illustrates the overall architecture of \qdkchem{}, highlighting the separation of client code, algorithm interfaces, factory registries, and data representations.

\subsection{Extensibility through plugins}

No single codebase can implement every method of interest.
Quantum chemistry has a rich ecosystem of specialized packages, and quantum computing frameworks continue to proliferate.
A useful toolkit must integrate with this ecosystem rather than stand apart from it.
Moreover, as quantum algorithms for chemistry evolve rapidly on both the hardware and algorithmic fronts, extensibility is essential for meeting current and future needs.

The plugin system in \qdkchem{} is a foundational design choice rather than an afterthought.
Treating extensibility as a first-class concern has become increasingly common in computational chemistry software, with major packages and workflow systems adopting modular or plugin-based architectures~\cite{Lehtola2023, DiFelice2023, Larsen2017, Sun2020, Smith2020Psi4, Huber2020, Richard2023, Barnes2024}.
In the quantum applications domain, similar extensibility patterns appear~\cite{McClean2020, McCaskey2020, Senicourt2022, Javadi-Abhari2024}.
In \qdkchem{}, plugins interact with the core exclusively through the same public interfaces available to any user, with no privileged access to internal state or implementation details.
This minimal coupling ensures that plugins can be developed, tested, and maintained independently of the core codebase and of each other, maximizing reuse potential and reducing the maintenance burden as both the toolkit and its extensions evolve.

\qdkchem{} uses a plugin architecture that supports two extension mechanisms.
Developers can add new implementations of existing algorithm types, enabling integration with external packages.
They can also define entirely new algorithm types with their own data contracts and factories, extending the workflow model itself.
This capability is distinct from merely extending an existing interface.

Once registered, plugin implementations appear in factory registries alongside native implementations and are discoverable at runtime.
Users need not distinguish between native and plugin-provided methods; the interface is uniform.
This symmetry is intentional: a plugin-provided algorithm is a first-class citizen, indistinguishable in use from a native implementation, yet developed and maintained without dependence on the internals of the core or other plugins.
This design allows \qdkchem{} to serve as a coordination layer for heterogeneous tools while preserving the independence of each component.

Alongside this plugin-based extensibility, \qdkchem{} includes robust, optimized native implementations for core workflow stages.
These implementations coexist in the same factory registries as plugin-provided methods, giving users flexibility to choose based on their needs: whether leveraging familiar external tools or using purpose-built implementations tuned for the toolkit's data structures.
The plugin architecture and native implementations are complementary.
Plugins extend the toolkit's reach into the broader ecosystem, while native implementations offer tightly integrated alternatives that have been developed and tested alongside the core infrastructure.

An important consequence of this architecture is its compatibility with commercial and proprietary software.
While the \qdkchem{} framework itself is permissively licensed and open source, the plugin infrastructure imposes no licensing requirements on extensions.
Plugins communicate with the core exclusively through public interfaces, without exposing implementation details in either direction.
This separation enables collaborators to develop plugins that wrap proprietary codes, protecting intellectual property and respecting licensing constraints.
Unlike ``push-in'' extensibility models that require contributions to reside within the main codebase, \qdkchem{} plugins can be maintained in separate repositories under whatever access controls their developers choose.
This design broadens the range of methods accessible through a unified interface, benefiting users who gain access to capabilities that might otherwise remain siloed.

\subsection{Multi-language support}

Quantum chemistry practitioners work in diverse environments: high-performance computing workflows often require compiled languages for efficiency, while rapid prototyping and interactive exploration favor Python.
\qdkchem{} addresses both contexts through a hybrid architecture: core data structures and algorithms are implemented in C++ for performance, with Python bindings exposing identical \ac{API} through \texttt{pybind11}~\cite{Wenzel2025}.
This approach follows successful precedents in packages such as PySCF~\cite{Sun2020} and Psi4~\cite{Smith2020Psi4}, which demonstrated that compiled computational kernels with Python-driven workflows can serve both high-performance and interactive use cases effectively.
Users can develop workflows in whichever language suits their context, and mixed-language pipelines are straightforward.

This design also accommodates diverse developer preferences and the broader ecosystem of compiled codes.
Researchers comfortable with Python can prototype new algorithms without learning C++, while performance-focused developers can work directly with the native implementation; both contribute through compatible interfaces.
Beyond C++ and Python, \qdkchem{} provides a consistent \ac{ABI} at plugin boundaries, enabling integration with external libraries written in C, Fortran, Rust, or other compiled languages through foreign function interfaces.
This approach meets developers where they are, rather than where a particular toolkit might expect them to be, and enables integration with binary-distributed libraries when source access is unavailable or impractical.

Multi-language support extends to quantum circuit specification as well.
Through the QDK, \qdkchem{} can target quantum algorithms expressed in Q\#, Cirq, Qiskit, and OpenQASM, allowing researchers to work in whichever quantum programming framework best suits their expertise or hardware requirements.
This flexibility ensures that users are not locked into a single quantum ecosystem and can leverage advances in any of these rapidly evolving frameworks.

\section{Capabilities}
\label{sec:capabilities}

\begin{figure}[t]
\centering
\resizebox{0.9\textwidth}{!}{%
\begin{tikzpicture}[
    row sep/.store in=\rowsep, row sep=4.5cm,
    box sep/.store in=\boxsep, box sep=0.5cm,
    %
    mainbox/.style={
        rectangle, rounded corners=8pt, 
        minimum width=6cm, text width=3.7cm, minimum height=1.5cm,
        text centered, font=\small\bfseries,
        text=white, align=center, line width=1pt,
        blur shadow={shadow blur steps=5, shadow xshift=0.5pt, shadow yshift=-1pt, shadow blur radius=2pt, shadow opacity=25}
    },
    classicalnode/.style={mainbox, fill=classicalbox, draw=classicaltext},
    quantumcircuitnode/.style={mainbox, fill=quantumcircuitbox, draw=quantumcircuittext},
    simnode/.style={mainbox, fill=simbox, draw=simtext},
    inputnode/.style={mainbox, fill=white, draw=inputborder, text=inputtext},
    postprocessnode/.style={mainbox, fill=postprocessbox, draw=postprocesstext},
    devtoolsubnode/.style={
        rectangle, rounded corners=6pt,
        minimum width=2.0cm, minimum height=1cm,
        text centered, font=\scriptsize\bfseries,
        text=white, fill=devtoolsub, draw=simbox, line width=1pt,
        blur shadow={shadow blur steps=5, shadow xshift=0.3pt, shadow yshift=-0.5pt, shadow blur radius=1.5pt, shadow opacity=20}
    },
    %
    clusterbox/.style={
        rectangle, rounded corners=12pt,
        line width=1.2pt, inner sep=12pt
    },
    %
    arrow/.style={
        -{Stealth[length=2.5mm, width=2mm, round]}, 
        line width=1.2pt, line cap=round, line join=round
    },
    classicalarrow/.style={arrow, classicaltext},
    quantumcircuitarrow/.style={arrow, quantumcircuittext},
    simarrow/.style={arrow, simtext},
    postprocessarrow/.style={arrow, postprocessbox},
]

\node[inputnode] (input) at (8, 0) {Input Structure};

\node[classicalnode] (activespace) at (8, -5) {Active Space Engineering};
\node[classicalnode, left=\boxsep of activespace] (scf) {Self-Consistent Field};
\node[classicalnode, left=\boxsep of scf] (geomopt) {Geometry \mbox{Optimization}};
\node[classicalnode, right=\boxsep of activespace] (hamilton) {Hamiltonian Formation};
\node[classicalnode, right=\boxsep of hamilton] (approxwfn) {Approximate Wavefunction};

\node[quantumcircuitnode] (pauliopt) at (8, -10) {Pauli Optimization};
\node[quantumcircuitnode, left=\boxsep of pauliopt] (hamencode) {Hamiltonian Encoding};
\node[quantumcircuitnode, right=\boxsep of pauliopt] (stateprep) {State Prep};

\node[simnode] (simulator) at (8, -15) {Simulator in QDK};
\node[simnode, left=5cm of simulator] (devtools) {Dev Tools};
\node[simnode, right=5cm of simulator] (quantumresults) {Quantum Results};
\node[simnode, below=1.5cm of simulator] (execute) {Quantum Hardware};

\node[devtoolsubnode, below=1.75cm of devtools, xshift=-2.25cm] (debug) {Debug};
\node[devtoolsubnode, right=\boxsep of debug] (visualization) {Visualization};
\node[devtoolsubnode, right=\boxsep of visualization] (copilot) {Copilot};
\node[devtoolsubnode, right=\boxsep of copilot] (compilation) {Compilation};

\node[postprocessnode] (postprocess) at (8, -22.75) {Processed Results};

\begin{scope}[on background layer]
    \node[clusterbox, fill=inputbg, draw=inputborder, fit=(input), inner sep=15pt,
          label={[font=\small\bfseries, inputtext]above:Chemical Structure Preparation}] (inputcluster) {};
    
    \node[clusterbox, fill=classicalbg, draw=classicalborder, fit=(geomopt)(approxwfn), inner sep=15pt,
          label={[font=\small\bfseries, classicaltext]above:Classical State Preparation}] (classicalcluster) {};
    
    \node[clusterbox, fill=quantumcircuitbg, draw=quantumcircuitborder, fit=(hamencode)(stateprep), inner sep=15pt,
          label={[font=\small\bfseries, quantumcircuittext]above:Building Quantum Circuit}] (qccluster) {};
    
    \node[clusterbox, fill=simbg, draw=simborder, fit=(devtools)(debug)(compilation)(execute)(quantumresults), inner sep=15pt,
          label={[font=\small\bfseries, simtext]above:Quantum Simulation and Execution}] (simcluster) {};
    
    \node[clusterbox, fill=postprocessbg, draw=postprocessborder, fit=(postprocess), inner sep=15pt,
          label={[font=\small\bfseries, postprocesstext]above:Classical Post Processing}] (postprocesscluster) {};
\end{scope}

\draw[classicalarrow] (geomopt) -- (scf);
\draw[classicalarrow] (scf) -- (activespace);
\draw[classicalarrow] (activespace) -- (hamilton);
\draw[classicalarrow] (hamilton) -- (approxwfn);

\draw[quantumcircuitarrow] (hamencode) -- (pauliopt);
\draw[quantumcircuitarrow] (pauliopt) -- (stateprep);

\foreach \angle/\target in {-110/debug, -90/visualization, -70/copilot, -50/compilation} {
    \draw[simarrow] (devtools.south) to[out=\angle, in=90, looseness=0.6] (\target.north);
}
\draw[simarrow] (devtools) to[out=0, in=180, looseness=0.5] (simulator);
\draw[simarrow] (devtools) to[out=-10, in=170, looseness=0.5] (execute);
\draw[simarrow] (simulator) -- (quantumresults);
\draw[simarrow] (execute) to[out=0, in=-135, looseness=0.5] (quantumresults.south west);

\draw[classicalarrow] (input) to[out=-90, in=90, looseness=0.6] (geomopt);
\draw[quantumcircuitarrow] (approxwfn) to[out=-90, in=90, looseness=0.6] (hamencode);
\draw[simarrow] (stateprep) to[out=-90, in=90, looseness=0.6] (devtools);
\draw[postprocessarrow] (quantumresults) to[out=-90, in=0, looseness=0.6] (postprocess.east);

\coordinate (feedback-bottom) at ($(geomopt.west)+(-1cm, -13cm)$);
\coordinate (feedback-mid) at ($(geomopt.west)+(-1cm, -8cm)$);
\draw[postprocessarrow, line width=1.5pt] 
    (postprocess.west) to[out=180, in=-90] (feedback-bottom)
                       to[out=90, in=-90] (feedback-mid)
                       to[out=90, in=180] (qccluster.west);

\end{tikzpicture}
}
\caption{An example workflow orchestrated using \qdkchem{}, from molecular geometry specification through to observable estimation.}
\label{fig:workflow}
\end{figure}
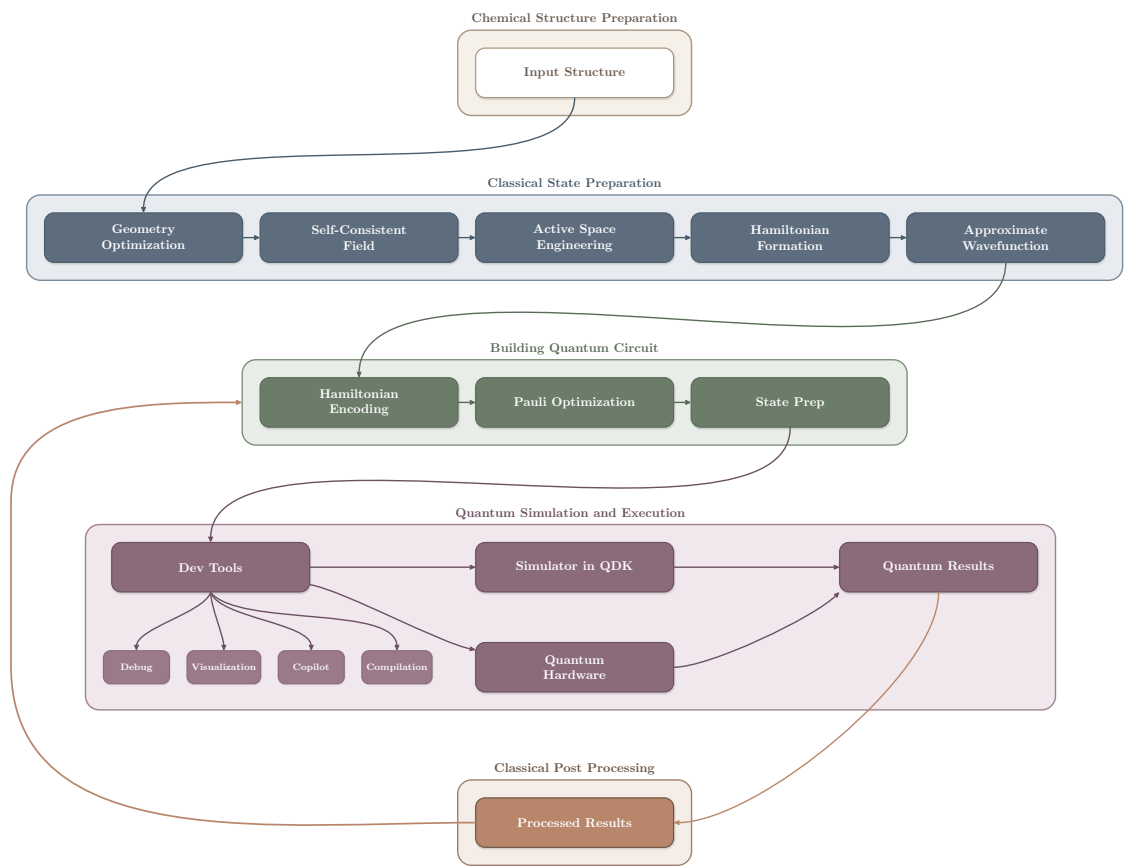

\qdkchem{} supports the stages of a typical quantum chemistry workflow, from molecular specification through observable estimation.
This section provides an overview of current capabilities; detailed documentation is available separately.
Figure~\ref{fig:workflow} illustrates the workflow stages, and Appendix~\ref{appendix:example-code} provides a complete code example.

\subsection{Classical electronic structure}
\label{sec:classical-electronic-structure}

Classical electronic structure methods provide the foundation for quantum chemistry workflows, generating the molecular orbitals, reference wavefunctions, and correlation treatments that serve as inputs for quantum algorithms.

\subsubsection{Molecular geometry handling}
\label{sec:geometry}

Typical quantum applications workflows begin with molecular geometry specification, which provides the atomic coordinates defining the system of interest. \qdkchem{} manages molecular geometries through the \texttt{Structure} class, which represents atomic coordinates as immutable, self-contained objects.
Currently, \qdkchem{} requires geometries to be provided as input; geometry optimization capabilities are planned for future releases.
Users can employ external packages for structure generation, e.g. geometry optimization, molecular dynamics, or experimental measurements, and import the resulting structures into \qdkchem{} for subsequent stages.
\qdkchem{} accepts geometries in standard formats including XYZ files~\cite{XYZ}, as well as toolkit-provided schemas for text (JSON) and binary (HDF5) serializations.
Molecular geometries arise from diverse sources---geometry optimization, reaction pathway sampling, molecular dynamics, experimental measurements, or manual construction---and \qdkchem{} accepts these geometries without prescribing their origin.
This design reflects the principle that structure generation is best handled by specialized tools, allowing users to employ their preferred methods for geometry preparation while leveraging \qdkchem{} for quantum chemistry workflow stages.
In larger pipelines, such as reaction network exploration but also molecular dynamics analysis, \qdkchem{} is intended as a final stage to generate the most accurate wavefunction and properties for important sampled structures.
This includes the analysis of a larger portion of sampled structures to determine their relevance by filtering based on energy criteria with e.g., \ac{DFT} and the analysis of optimal active space sizes to decide if calculations require a quantum computer.

\subsubsection{Self-consistent field calculations}
\label{sec:scf}

Mean-field \ac{SCF} calculations produce initial wavefunctions and reference energies that anchor subsequent correlation treatments.
The native \ac{SCF} solver implements the geometric direct minimization algorithm~\cite{VanVoorhis2002}, which frames orbital optimization as energy minimization on a Grassmann manifold rather than iterative diagonalization.
This formulation provides robust convergence for systems where conventional diagonalization-based approaches struggle, including open-shell configurations, small-gap molecules, and transition metal complexes---precisely the systems where quantum algorithms are expected to provide the most significant improvements over classical methods~\cite{Reiher2017,Cao2019,Lee2021}.
Both \ac{HF} and \ac{DFT} methods are supported, with restricted, unrestricted, and restricted open-shell variants available for each.
The native implementation builds on established libraries: Libint~\cite{Libint2} for molecular integrals, Libecpint~\cite{Shaw2017,Shaw2021} for effective core potential integrals, GauXC~\cite{Petrone2018,williams20on,williams2021achieving,williams2023distributed} for numerical integration on atom-centered grids, and Libxc~\cite{Lehtola2018} for exchange-correlation functional evaluation.
In addition, the PySCF~\cite{Sun2020} plugin provides access to its comprehensive suite of electronic structure methods through the uniform \qdkchem{} interface, enabling users to leverage this familiar tool within the modular framework.

Challenging \ac{SCF} problems frequently converge to saddle points rather than true minima, producing reference states that compromise the quality of subsequent many-body calculations.
\qdkchem{} provides stability analysis tools~\cite{Schlegel1991} to detect these problematic solutions by examining the Hessian of the energy with respect to orbital rotations.
When an instability is identified, users can perturb the wavefunction along the unstable mode and reoptimize using any implementation of the self-consistent field solver.
This modular approach---separating stability detection from reoptimization---allows users to inspect intermediate results, apply domain-specific adjustments, and combine stability analysis with other orbital manipulation techniques as their workflow requires.

\subsubsection{Orbital localization}
\label{sec:orbital-localization}

Orbital localization transforms delocalized canonical orbitals into more compact representations that improve chemical interpretability, accelerate correlation method convergence, and facilitate active space selection.
Canonical orbitals from \ac{SCF} calculations are typically delocalized across the entire molecule, which can obscure chemical interpretation and slow the convergence of post-mean-field correlation methods.
\qdkchem{} provides several classes of orbital transformation techniques to yield representations that are ``localized'' in various senses: spatially compact, ordered by correlation importance, or partitioned by entanglement character.
Each of these localization behaviors is beneficial for different downstream applications.

Spatial localization methods produce orbitals concentrated on bonds, lone pairs, or at\-omic sites by iteratively optimizing a cost function.
\qdkchem{} supports several established approaches through native implementations and plugin integrations, including through PySCF~\cite{Sun2020}: Pipek--Mezey~\cite{Pipek1989} localization (minimizing orbital spread via Mulliken populations), Foster--Boys~\cite{Foster1960} localization (minimizing orbital second moments), and Edmiston--Ruedenberg~\cite{Edmiston1963} localization (maximizing self-repulsion integrals), each yielding orbitals with distinct characteristics~\cite{Boughton1993,Hoyvik2012a,Hoyvik2012b,Lehtola2013}.

Beyond spatial localization, orbital transformations can organize orbitals by correlation character.
Natural orbitals~\cite{Lowdin1956}, obtained by diagonalizing the one-particle reduced density matrix, order orbitals by occupation number; fractional occupations indicate significant correlation effects.
\qdkchem{} provides a native implementation of \ac{MP2} natural orbitals~\cite{Pulay1986,Bofill1989}, offering an efficient route to correlation-ordered orbitals without full multi-configuration calculations.
This perspective complements spatial localization: spatially localized orbitals aid chemical interpretation and basis truncation, while natural orbital ordering guides active space selection and quantum resource allocation.

Localization in near-complete basis sets is numerically ill-posed for most iterative methods.
\qdkchem{} addresses this through the valence-virtual/hard-virtual separation~\cite{Subotnik2005,Wang2025}, partitioning virtual orbitals into chemically relevant and numerically problematic subspaces.
This yields orbitals that vary smoothly with molecular geometry, which is valuable for consistent active space selection along reaction coordinates.

\subsubsection{Active space selection}
\label{sec:active-space}

Active space selection identifies the subset of molecular orbitals that exhibit significant correlation effects and must be treated explicitly on the quantum computer, while freezing the remaining orbitals into classical contributions.
Reducing quantum resource requirements is essential for early fault-tolerant quantum computers, and active space selection is among the most effective strategies for achieving this reduction.
The challenge for active space selection is identifying which molecular orbitals exhibit significant correlation character and must be treated explicitly, while freezing the remaining orbitals into effective core and virtual subspaces that contribute only classically.
An optimal active space should capture the essential many-body physics while remaining compact enough to fit within hardware qubit and gate budgets.

\qdkchem{} provides both automated and manual approaches to active space selection.
Entropy-based methods leverage concepts from quantum information theory to automatically identify strongly correlated orbitals based on their entanglement with the rest of the system~\cite{Legeza2003,Legeza2006,Rissler2006,Boguslawski2015}.
A native implementation of the autoCAS algorithm~\cite{Stein2016,Stein2019} computes single-orbital entropies from the one- and two-electron reduced density matrices of a multi-configurational wavefunction.
Orbitals with high entropy---indicating strong entanglement with the remainder of the system---are included in the active space, while weakly entangled orbitals are frozen.
The implementation includes both standard autoCAS and an enhanced variant (autoCAS-EOS) for improved robustness in challenging cases~\cite{AutoCASEOS}.

Occupation-based methods provide automatic selection using natural orbital occupation numbers from correlated many-body methods.
Orbitals with fractional occupations indicate incomplete filling and associated correlation effects, warranting inclusion in the active space~\cite{Bofill1989,Pulay1988,Khedkar2019}.
The \ac{AVAS}~\cite{Sayfutyarova2017} method, available through the PySCF plugin, projects molecular orbitals onto a target atomic orbital basis---for example, metal 3d orbitals---to systematically identify valence active spaces appropriate for transition metal chemistry.

Valence-based selection offers a simpler approach, specifying active electrons and orbitals centered around the Fermi level based on chemical intuition about frontier orbital importance.
Beyond these structured approaches, \qdkchem{} supports fully manual active space specification by directly populating the active space data within orbital objects.
This capability enables expert users with domain-specific knowledge or superior chemical intuition to inject custom orbital partitionings that may not arise from automated selection criteria, ensuring that the toolkit accommodates both routine applications and specialized research workflows.

\subsubsection{Multi-configuration wavefunctions}
\label{sec:mc-wavefunction}

With an active space Hamiltonian constructed, \ac{MC} calculations produce wavefunctions that capture correlation effects beyond the single-determinant mean-field approximation.
While these methods provide more accurate energy estimates than \ac{SCF} calculations, their primary role in quantum application workflows is twofold: generating high-quality initial states for quantum algorithms, and providing classical baselines against which quantum results can be compared.
On scaled fault-tolerant quantum computers, classically computed \ac{MC} wavefunctions serve as the foundation for state preparation circuits, enabling algorithms such as quantum phase estimation to target chemical accuracy for systems where classical methods face fundamental scaling limitations.

\qdkchem{} supports two main classes of \ac{MC} methods: orbital-frozen configuration interaction methods and orbital-optimized \ac{MCSCF} methods.
Both native \ac{CASCI}/\ac{FCI} solvers and access to the comprehensive \ac{CAS} capabilities of PySCF~\cite{Sun2020} are available through the uniform \qdkchem{} interface, providing users with multiple options for classical active space correlation treatments.
\Ac{CASCI} calculates an exact solution within the defined active space, with core orbitals frozen and virtual orbitals excluded~\cite{Roos1980a}.
However, the exponential scaling of exact diagonalization limits \ac{CASCI} to modest active spaces.

Selected configuration interaction methods approximate the exact solution by iteratively identifying and including only the most important configurations, enabling treatment of substantially larger active spaces at the cost of controlled approximation~\cite{Huron1973,Holmes2016,Sharma2017,Zhang2025TrimCI}.
\qdkchem{} employs the \ac{ASCI} variant~\cite{Tubman2016,Tubman2020} through native integration with \ac{MACIS}~\cite{Williams-Young2023}, a high-performance selected configuration library for large-scale calculations.
\ac{ASCI} iteratively grows the determinant space by identifying configurations with the largest contributions to the wavefunction, achieving near-\ac{CASCI} accuracy at a fraction of the computational cost and enabling treatment of active spaces that would otherwise be intractable.
For users who prefer alternative multi-configuration implementations or require additional correlation treatments, the PySCF~\cite{Sun2020} plugin provides access to its wide range of multi-configuration and correlation methods through the same uniform interface.

The \ac{MC} methods described in this section are entirely classical---they execute on conventional hardware and produce deterministic outputs including wavefunction coefficients and reduced density matrices.
In this sense, classical \ac{CASCI} and quantum \ac{CASCI} solvers address the same underlying problem but differ in execution model and output contracts.
As illustrated in Figure~\ref{fig:workflow}, the ``Classical State Preparation'' phase encompasses these classical correlation methods, while the subsequent ``Building Quantum Circuit'' and ``Quantum Simulation and Execution'' phases represent the transition to quantum computation.
Within \qdkchem{}, quantum algorithms serve as an alternative means of solving the active space problem---one that may ultimately scale to system sizes intractable for classical methods, but differ fundamentally in the nature of their outputs and execution requirements.
We discuss quantum algorithms for energy estimation in Sections~\ref{sec:energy-estimation} and~\ref{sec:qpe}.

\subsection{Quantum circuit construction}
\label{sec:quantum-circuit}

Quantum circuit construction transforms classical electronic structure data into executable quantum programs, encompassing the encoding of molecular Hamiltonians onto qubits, preparation of initial quantum states, and extraction of observable estimates through measurement.

\subsubsection{Qubit Hamiltonian construction}
\label{sec:qubit-hamiltonians}

Qubit Hamiltonian construction transforms fermionic molecular Hamiltonians into qubit operator representations suitable for quantum computation, encoding the electronic structure problem in a form that quantum hardware can process.
Classical quantum chemistry methods express electronic interactions using second quantization, with fermionic creation and annihilation operators describing electron interactions.
Quantum computers operate on qubits, necessitating a transformation from fermionic operators to qubit operators that preserves the anti-commutation relations essential to fermionic statistics.
Standard fermion-to-qubit mappings include the Jordan--Wigner~\cite{Jordan-Wigner1928}, Bravyi--Kitaev~\cite{Bravyi-Kitaev2002}, and parity~\cite{Love2012} transformations, each offering different trade-offs in operator weight and qubit connectivity requirements.
More recent developments have produced mappings optimized for particular hardware topologies or molecular symmetries~\cite{Setia2019,Steudtner2019,Derby2021,Miller2023,Liu2025}.

\qdkchem{} implements the standard fermion-to-qubit transformations, representing the resulting qubit Hamiltonian as a sum of Pauli strings with associated coefficients.
Custom encodings, including hardware-aware and system-adapted mappings, can be integrated through the plugin system.

\subsubsection{State preparation}
\label{sec:state-preparation}

Quantum algorithms for chemistry require preparing quantum states that approximate the ground or excited states of molecular systems.
In \qdkchem{}, state preparation is viewed as a mapping between a classical representation of the molecular wavefunction---a Slater determinant or linear combination thereof, represented by the \texttt{Wavefunction} class---and a quantum circuit that prepares the corresponding state on a quantum computer given a particular qubit encoding.

This mapping can be accomplished through various techniques, including isometry encoding~\cite{Christandl2016} for general states and linear combinations of unitaries for structured preparations~\cite{Childs2012,Berry2015,Low2019,Gilyen2019,Chakraborty2024,Sanders2019,Fomichev2024}.
However, when wavefunctions are sparse---containing few configurations with significant amplitudes---generic methods can be inefficient.
\qdkchem{} provides a specialized method for sparse wavefunction state preparation based on sparse isometries~\cite{Malvetti2021, SparseIsometry}, which avoid the exponential scaling associated with generic methods.

State preparation in \qdkchem{} is modular and decoupled from the choice of downstream quantum algorithm.
Users can inject their own state preparation circuits without modifying any other part of the workflow.
The same state preparation circuit can feed into statistical energy estimation, quantum phase estimation, or other downstream algorithms, with the choice of preparation method orthogonal to the choice of observable extraction strategy.
This separation reflects a practical reality: optimal state preparation depends on both the molecular system and the available quantum resources, and researchers developing novel preparation techniques benefit from the ability to integrate their methods into a complete workflow without reconstructing the surrounding infrastructure. 

\subsubsection{Statistical energy estimation}
\label{sec:energy-estimation}

One approach to estimating observable expectation values, such as ground state energies, is through statistical sampling of measurements performed on prepared quantum states.
This approach forms the foundation of \ac{VQE} algorithms~\cite{Peruzzo2014,McClean2016,Kandala2017},
where energy estimates guide classical optimization of parameterized quantum circuits.
More broadly, statistical sampling applies whenever information must be extracted from quantum states, whether for variational optimization, benchmarking prepared states against classical references, validating quantum algorithm outputs, or reconstructing quantum states through tomographic techniques~\cite{Huang2020,Korhonen2025,Korhonen2025b,Mangini2025}.
\qdkchem{} supports this post-processing procedure through native implementations and plugin integrations.

This procedure begins by decomposing an operator of interest into a set of measurable components.
For example, starting from a qubit-mapped Hamiltonian, represented as a linear combination of Pauli strings, terms can be grouped into sets of mutually commuting operators that share measurement bases, reducing the number of distinct circuit executions required~\cite{Gokhale2019,Yen2020,Verteletskyi2020}.
\qdkchem{} provides qubit-wise commutativity grouping through its Qiskit~\cite{Javadi-Abhari2024} plugin, enabling efficient batching of measurements and access to Qiskit's circuit construction and transpilation capabilities.
Additional optimizations filter Pauli terms with negligible expectation values given the prepared state, pre-computing their classical contributions rather than consuming quantum resources.

Measurement circuits execute on quantum hardware or simulators to obtain statistical samples for each operator group.
Classical post-processing aggregates these measurement outcomes, combining statistics from each group with their corresponding Hamiltonian coefficients and pre-computed classical contributions to yield energy estimates.
The statistical nature of quantum measurements introduces variance that decreases with increasing shot count, and \qdkchem{} provides utilities for reassembling these measurements into final expectation values with uncertainty quantification (e.g., variance estimation).

\subsubsection{Quantum phase estimation}
\label{sec:qpe}

While statistical energy estimation provides a practical approach for near-term devices, \ac{QPE} offers a fundamentally different paradigm for extracting eigenvalues from quantum systems~\cite{Kitaev1995,Abrams1999,Nielsen-Chuang2010-QPE}.
In the context of active space methods, \ac{QPE} can be viewed as a quantum \ac{CASCI} solver: it targets the same electronic structure problem within the defined active space, but executes on quantum hardware rather than classical processors.
The core structure of \ac{QPE} consists of controlled unitary operations applied to a trial state, followed by an inverse quantum Fourier transform that extracts the encoded spectral information into a computational basis measurement.
This modular structure naturally decomposes into independent components: state preparation, controlled unitary implementation, and phase readout.

The design space for each component admits significant variation.
State preparation techniques range from simple product states to sophisticated multi-configuration ansätze.
Controlled unitary implementation---the computational core of \ac{QPE}---encompasses multiple paradigms: product formula approaches such as Suzuki-Trotter decomposition~\cite{Hatano2005} approximate time evolution through sequences of simpler exponentials, while block encoding techniques~\cite{Low2017,Low2019,Berry2019,Gilyen2019} embed the Hamiltonian into a larger unitary matrix, enabling quantum signal processing and qubitization methods with improved asymptotic scaling.
Phase readout variants include standard \ac{QPE}~\cite{Kitaev1995, Nielsen-Chuang2010-QPE} using multiple ancilla qubits for parallel phase extraction, and iterative approaches~\cite{Dobsicek2007} that trade parallelism for reduced qubit overhead through sequential single-ancilla measurements.
These algorithmic choices interact with hardware constraints: statistical estimation tolerates shorter coherence times suitable for near-term devices, while QPE achieves precision through circuit depth rather than shot count, becoming advantageous when hardware can sustain the required coherent evolution~\cite{Dobsicek2007,OBrien2019,Dutkiewicz2025,Berry2015,Babbush2018}.

The \qdkchem{} implementation of \ac{QPE} exemplifies the toolkit's modular composability, factoring the algorithm into distinct layers that mirror this design space.
The unitary construction layer currently provides product formula decompositions, with the interface designed to accommodate block encoding implementations as they become available through native development or plugin contributions.
The mapper layer synthesizes constructed operators into controlled unitary circuits, while the phase extraction layer supports both standard and iterative readout variants.
Each layer exposes its own algorithm interface through the standard factory-based extensibility mechanism, enabling researchers to contribute novel techniques---whether improved Hamiltonian simulation methods, optimized circuit synthesis strategies, or alternative phase extraction protocols---while relying on validated implementations elsewhere.

\ac{QPE} execution uses the component specifications as input parameters that are composed internally into the complete circuit.
This design enables transparent testing of algorithmic innovations: a new technique can be validated using \ac{QDK}- or Qiskit-derived simulators before deployment to hardware providers, with the same specification producing identical circuit structure regardless of backend.
For quantum circuit construction and execution, \qdkchem{} offers parallel pathways through different backend ecosystems.
The \ac{QDK}~\cite{QDK} integration provides access to Q\#-based algorithm development, \ac{QDK}'s optimizing compilers, and its suite of high-performance simulators---including sparse state-vector, full state-vector, and resource estimation backends.
The Qiskit~\cite{Javadi-Abhari2024} plugin provides an alternative pathway with circuit construction, transpilation, and execution capabilities targeting other Qiskit-compatible backends.
Users can choose whichever ecosystem best suits their needs, and the uniform \qdkchem{} interface ensures that workflow logic remains unchanged regardless of backend selection, allowing the path from prototyping to hardware execution without workflow modifications.

\section{Conclusions}
\label{sec:conclusions}

\qdkchem{} provides infrastructure for quantum chemistry workflows that bridges classical electronic structure and quantum algorithm design.
Beyond delivering high-performance software components, the toolkit's architecture enables diverse methods to interoperate within a unified framework by separating data from algorithms, exposing factory-based interfaces, and supporting plugins for external packages.

This design serves the varied needs of the quantum chemistry community.
Method developers can contribute new algorithms at any workflow stage while relying on validated implementations elsewhere.
Application scientists can systematically compare approaches by swapping components through factory interfaces rather than rewriting pipeline logic.
As quantum hardware matures, individual workflow stages can evolve independently---incorporating new methods, hardware interfaces, or error mitigation strategies---without disrupting established workflows.
Serialization support for all intermediate artifacts provides the provenance tracking essential for reproducible science.

We view \qdkchem{} as a foundation for community development.
The source code is available at \url{https://github.com/microsoft/qdk-chemistry}, and the package can be installed via \texttt{pip install qdk-chemistry} from PyPI.\footnote{\url{https://pypi.org/project/qdk-chemistry/}}
Researchers can develop and distribute extensions independently, without requiring changes to the core codebase or coordination with other contributors.
By reducing the friction of building, sharing, and comparing quantum chemistry workflows, we aim to accelerate progress in understanding which classical and quantum methods best address the electronic structure problems that motivate this field.

\section*{Acknowledgements}

The authors thank
Markus Reiher,
Valentin Barandun,
Matthias Christandl,
Karol Kowalski,
and the Algorithmiq team
for helpful conversations and/or feedback on \qdkchem{}.

\bibliographystyle{unsrtnat}
\bibliography{references}

@misc{XYZ,
    title = {{XYZ} file format},
    howpublished = {Open Babel documentation},
    url = {https://openbabel.org/docs/FileFormats/XYZ_cartesian_coordinates_format.html},
}

@article{Korhonen2025,
    author = {Korhonen, Keijo and Vappula, Hetta and Glos, Adam and Cattaneo, Marco and Zimbor\'as, Zolt\'an and Borrelli, Elsi-Mari and Rossi, Matteo A. C.},
    title = {Practical techniques for high-precision measurements on near-term quantum hardware and applications in molecular energy estimation},
    journal = {npj Quantum Information},
    volume = {11},
    pages = {110},
    year = {2025},
    doi = {10.1038/s41534-025-01066-1}
}

@article{Korhonen2025b,
    author = {Korhonen, Keijo and Mangini, Stefano and Malmi, Joonas and Vappula, Hetta and Cavalcanti, Daniel},
    title = {Improving shadow estimation with locally-optimal dual frames},
    year = {2025},
    eprint = {2511.02555},
    archivePrefix = {arXiv},
    primaryClass = {quant-ph},
    doi = {10.48550/arXiv.2511.02555}
}

@article{Mangini2025,
    author = {Mangini, Stefano and Cavalcanti, Daniel},
    title = {Low variance estimations of many observables with tensor networks and informationally-complete measurements},
    journal = {Quantum},
    volume = {9},
    pages = {1812},
    year = {2025},
    doi = {10.22331/q-2025-07-23-1812}
}

@article{Reiher2017,
    author = {Markus Reiher and Nathan Wiebe and Krysta M. Svore and Dave Wecker and Matthias Troyer},
    title = {Elucidating reaction mechanisms on quantum computers},
    journal = {Proceedings of the National Academy of Sciences},
    volume = {114},
    number = {29},
    pages = {7555-7560},
    year = {2017},
    doi = {10.1073/pnas.1619152114},
}

@article{Lee2021,
  title = {Even More Efficient Quantum Computations of Chemistry Through Tensor Hypercontraction},
  author = {Lee, Joonho and Berry, Dominic W. and Gidney, Craig and Huggins, William J. and McClean, Jarrod R. and Wiebe, Nathan and Babbush, Ryan},
  journal = {PRX Quantum},
  volume = {2},
  year={2021},
  issue = {3},
  pages = {030305},
  doi = {10.1103/PRXQuantum.2.030305},
}

@article{Cao2019,
author = {Cao, Yudong and Romero, Jonathan and Olson, Jonathan P. and Degroote, Matthias and Johnson, Peter D. and Kieferová, Mária and Kivlichan, Ian D. and Menke, Tim and Peropadre, Borja and Sawaya, Nicolas P. D. and Sim, Sukin and Veis, Libor and Aspuru-Guzik, Alán},
title = {Quantum Chemistry in the Age of Quantum Computing},
journal = {Chemical Reviews},
volume = {119},
number = {19},
pages = {10856-10915},
year = {2019},
doi = {10.1021/acs.chemrev.8b00803},
}

@article{Hoyvik2012a,
    author = {Høyvik, Ida-Marie and Jansik, Branislav and Jørgensen, Poul},
    title = {Orbital localization using fourth central moment minimization},
    journal = {The Journal of Chemical Physics},
    volume = {137},
    number = {22},
    pages = {224114},
    year = {2012},
    doi = {10.1063/1.4769866},
}

@article{Hoyvik2012b,
author = {Høyvik, Ida-Marie and Jansik, Branislav and Jørgensen, Poul},
title = {Trust Region Minimization of Orbital Localization Functions},
journal = {Journal of Chemical Theory and Computation},
volume = {8},
number = {9},
pages = {3137-3146},
year = {2012},
doi = {10.1021/ct300473g},
}

@article{Pulay1986,
author={Peter Pulay and Svein Saebø},
title={Orbital-invariant formulation and second-order gradient evaluation in {Møller-Plesset} perturbation theory},
journal={Theoretica chimica acta},
volume={69},
pages={357-368},
year={1986},
doi={10.1007/BF00526697},
}

@article{Bofill1989,
    author = {Bofill, Josep M. and Pulay, Peter},
    title = {The unrestricted natural orbital-complete active space {(UNO-CAS)} method: An inexpensive alternative to the complete active space-self-consistent-field {(CAS-SCF)} method},
    journal = {The Journal of Chemical Physics},
    volume = {90},
    number = {7},
    pages = {3637-3646},
    year = {1989},
    doi = {10.1063/1.455822},
}

@article{Legeza2003,
  title = {Optimizing the density-matrix renormalization group method using quantum information entropy},
  author = {Legeza, \"O. and S\'olyom, J.},
  journal = {Phys. Rev. B},
  volume = {68},
  issue = {19},
  pages = {195116},
  year = {2003},
  doi = {10.1103/PhysRevB.68.195116},
}

@article{Legeza2006,
  title = {Two-Site Entropy and Quantum Phase Transitions in Low-Dimensional Models},
  author = {Legeza, \"O. and S\'olyom, J.},
  journal = {Phys. Rev. Lett.},
  volume = {96},
  issue = {11},
  pages = {116401},
  year = {2006},
  doi = {10.1103/PhysRevLett.96.116401},
}

@article{Rissler2006,
author = {Jörg Rissler and Reinhard M. Noack and Steven R. White},
title = {Measuring orbital interaction using quantum information theory},
journal = {Chemical Physics},
volume = {323},
number = {2},
pages = {519-531},
year = {2006},
doi = {10.1016/j.chemphys.2005.10.018},
}

@article{Stein2016,
author = {Stein, Christopher J. and Reiher, Markus},
title = {Automated Selection of Active Orbital Spaces},
journal = {Journal of Chemical Theory and Computation},
volume = {12},
number = {4},
pages = {1760-1771},
year = {2016},
doi = {10.1021/acs.jctc.6b00156},
}

@article{Berry2019,
  doi = {10.22331/q-2019-12-02-208},
  title = {Qubitization of Arbitrary Basis Quantum Chemistry Leveraging Sparsity and Low Rank Factorization},
  author = {Berry, Dominic W. and Gidney, Craig and Motta, Mario and McClean, Jarrod R. and Babbush, Ryan},
  journal = {{Quantum}},
  volume = {3},
  pages = {208},
  year = {2019}
}

@article{Huron1973,
    author = {Huron, B. and Malrieu, J. P. and Rancurel, P.},
    title = {Iterative perturbation calculations of ground and excited state energies from multiconfigurational zeroth-order wavefunctions},
    journal = {The Journal of Chemical Physics},
    volume = {58},
    number = {12},
    pages = {5745-5759},
    year = {1973},
    doi = {10.1063/1.1679199},
}

@article{Holmes2016,
author = {Holmes, Adam A. and Tubman, Norm M. and Umrigar, C. J.},
title = {Heat-Bath Configuration Interaction: An Efficient Selected Configuration Interaction Algorithm Inspired by Heat-Bath Sampling},
journal = {Journal of Chemical Theory and Computation},
volume = {12},
number = {8},
pages = {3674-3680},
year = {2016},
doi = {10.1021/acs.jctc.6b00407},
}

@article{Sharma2017,
author = {Sharma, Sandeep and Holmes, Adam A. and Jeanmairet, Guillaume and Alavi, Ali and Umrigar, C. J.},
title = {Semistochastic Heat-Bath Configuration Interaction Method: Selected Configuration Interaction with Semistochastic Perturbation Theory},
journal = {Journal of Chemical Theory and Computation},
volume = {13},
number = {4},
pages = {1595-1604},
year = {2017},
doi = {10.1021/acs.jctc.6b01028},
}

@article{Roos1980a,
title = {A complete active space {SCF} method ({CASSCF}) using a density matrix formulated super-{CI} approach},
journal = {Chemical Physics},
volume = {48},
number = {2},
pages = {157-173},
year = {1980},
doi = {10.1016/0301-0104(80)80045-0},
author = {Björn O. Roos and Peter R. Taylor and Per E.M. Siegbahn},
}

@article{Derby2021,
  title = {Compact fermion to qubit mappings},
  author = {Derby, Charles and Klassen, Joel and Bausch, Johannes and Cubitt, Toby},
  journal = {Phys. Rev. B},
  volume = {104},
  issue = {3},
  pages = {035118},
  year = {2021},
  doi = {10.1103/PhysRevB.104.035118},
}

@article{Setia2019,
  title = {Superfast encodings for fermionic quantum simulation},
  author = {Setia, Kanav and Bravyi, Sergey and Mezzacapo, Antonio and Whitfield, James D.},
  journal = {Phys. Rev. Res.},
  volume = {1},
  issue = {3},
  pages = {033033},
  year = {2019},
  doi = {10.1103/PhysRevResearch.1.033033},
}

@article{Steudtner2019,
  title = {Quantum codes for quantum simulation of fermions on a square lattice of qubits},
  author = {Steudtner, Mark and Wehner, Stephanie},
  journal = {Phys. Rev. A},
  volume = {99},
  issue = {2},
  pages = {022308},
  year = {2019},
  doi = {10.1103/PhysRevA.99.022308},
}

@article{Miller2023,
  title = {Bonsai Algorithm: Grow Your Own Fermion-to-Qubit Mappings},
  author = {Miller, Aaron and Zimbor\'as, Zolt\'an and Knecht, Stefan and Maniscalco, Sabrina and Garc\'{\i}a-P\'erez, Guillermo},
  journal = {PRX Quantum},
  volume = {4},
  issue = {3},
  pages = {030314},
  year = {2023},
  publisher = {American Physical Society},
  doi = {10.1103/PRXQuantum.4.030314},
}

@article{Peruzzo2014,
  title={A variational eigenvalue solver on a photonic quantum processor},
  author={Peruzzo, Alberto and McClean, Jarrod and Shadbolt, Peter and Yung, Man-Hong and Zhou, Xiao-Qi and Love, Peter J. and Aspuru-Guzik, Al{\'a}n and {O'Brien}, Jeremy L.},
  journal={Nature Communications},
  volume={5},
  number={1},
  pages={4213},
  year={2014},
  doi={10.1038/ncomms5213}
}

@article{McClean2016,
doi = {10.1088/1367-2630/18/2/023023},
year = {2016},
volume = {18},
number = {2},
pages = {023023},
author = {McClean, Jarrod R and Romero, Jonathan and Babbush, Ryan and Aspuru-Guzik, Alán},
title = {The theory of variational hybrid quantum-classical algorithms},
journal = {New Journal of Physics},
}

@article{Kandala2017,
  title={Hardware-efficient variational quantum eigensolver for small molecules and quantum magnets},
  author={Kandala, Abhinav and Mezzacapo, Antonio and Temme, Kristan and Takita, Maika and Brink, Markus and Chow, Jerry M and Gambetta, Jay M},
  journal={Nature},
  volume={549},
  number={7671},
  pages={242--246},
  year={2017},
  doi={10.1038/nature23879},
}

@article{Huang2020,
  title={Predicting many properties of a quantum system from very few measurements},
  author={Huang, Hsin-Yuan and Kueng, Richard and Preskill, John},
  journal={Nature Physics},
  volume={16},
  number={10},
  pages={1050--1057},
  year={2020},
  doi={10.1038/s41567-020-0932-7}
}

@article{Gokhale2019,
      title={Minimizing State Preparations in Variational Quantum Eigensolver by Partitioning into Commuting Families},
      author={Pranav Gokhale and Olivia Angiuli and Yongshan Ding and Kaiwen Gui and Teague Tomesh and Martin Suchara and Margaret Martonosi and Frederic T. Chong},
      year={2019},
      journal={arXiv},
      url={https://arxiv.org/abs/1907.13623},
}

@article{Yen2020,
      title={Measuring all compatible operators in one series of a single-qubit measurements using unitary transformations},
      author={Tzu-Ching Yen and Vladyslav Verteletskyi and Artur F. Izmaylov},
      year={2020},
      journal={arXiv},
      url={https://arxiv.org/abs/1907.09386},
}

@article{Verteletskyi2020,
    author = {Verteletskyi, Vladyslav and Yen, Tzu-Ching and Izmaylov, Artur F.},
    title = {Measurement optimization in the variational quantum eigensolver using a minimum clique cover},
    journal = {The Journal of Chemical Physics},
    volume = {152},
    number = {12},
    pages = {124114},
    year = {2020},
    doi = {10.1063/1.5141458},
}

@article{Abrams1999,
  title = {Quantum Algorithm Providing Exponential Speed Increase for Finding Eigenvalues and Eigenvectors},
  author = {Abrams, Daniel S. and Lloyd, Seth},
  journal = {Phys. Rev. Lett.},
  volume = {83},
  issue = {24},
  pages = {5162--5165},
  year = {1999},
  doi = {10.1103/PhysRevLett.83.5162},
}

@article{Berry2015,
  title = {Simulating {Hamiltonian} Dynamics with a Truncated {Taylor} Series},
  author = {Berry, Dominic W. and Childs, Andrew M. and Cleve, Richard and Kothari, Robin and Somma, Rolando D.},
  journal = {Phys. Rev. Lett.},
  volume = {114},
  issue = {9},
  pages = {090502},
  year = {2015},
  doi = {10.1103/PhysRevLett.114.090502},
}

@article{Low2017,
  title = {Optimal {Hamiltonian} Simulation by Quantum Signal Processing},
  author = {Low, Guang Hao and Chuang, Isaac L.},
  journal = {Phys. Rev. Lett.},
  volume = {118},
  issue = {1},
  pages = {010501},
  year = {2017},
  doi = {10.1103/PhysRevLett.118.010501},
}

@article{Low2019,
  title = {Hamiltonian Simulation by Qubitization},
  author = {Low, Guang Hao and Chuang, Isaac L.},
  journal = {Quantum},
  volume = {3},
  pages = {163},
  year = {2019},
  doi = {10.22331/q-2019-07-12-163},
}

@article{Gilyen2019,
  title = {Quantum singular value transformation and beyond: exponential improvements for quantum matrix arithmetics},
  author = {Gily\'en, Andr\'as and Su, Yuan and Low, Guang Hao and Wiebe, Nathan},
  journal = {Proceedings of the 51st Annual ACM SIGACT Symposium on Theory of Computing},
  pages = {193--204},
  year = {2019},
  doi = {10.1145/3313276.3316366},
}

@article{Babbush2018,
  title = {Encoding Electronic Spectra in Quantum Circuits with Linear {T} Complexity},
  author = {Babbush, Ryan and Gidney, Craig and Berry, Dominic W. and Wiebe, Nathan and McClean, Jarrod and Paler, Alexandru and Fowler, Austin and Neven, Hartmut},
  journal = {Phys. Rev. X},
  volume = {8},
  issue = {4},
  pages = {041015},
  year = {2018},
  doi = {10.1103/PhysRevX.8.041015},
}

@article{OBrien2019,
doi = {10.1088/1367-2630/aafb8e},
year = {2019},
volume = {21},
number = {2},
pages = {023022},
author = {O’Brien, Thomas E and Tarasinski, Brian and Terhal, Barbara M},
title = {Quantum phase estimation of multiple eigenvalues for small-scale (noisy) experiments},
journal = {New Journal of Physics},
}

@article{Sayfutyarova2017,
    title={Automated construction of molecular active spaces from atomic valence orbitals},
    author={Elvira R. Sayfutyarova and Qiming Sun and Garnet Kin-Lic Chan and Gerald Knizia},
    journal={Journal of Chemical Theory and Computation},
    volume={13},
    number={9},
    pages={4063-4078},
    year={2017},
    doi={10.1021/acs.jctc.7b00128},
}

@article{Boughton1993,
    title={Comparison of the {Boys} and {Pipek-Mezey} localizations in the local correlation approach and automatic virtual basis selection},
    author={James W. Boughton and Peter Pulay},
    journal={Journal of Computational Chemistry},
    volume={14},
    number={6},
    pages={736-740},
    year={1993},
    doi={10.1002/jcc.540140615}
}

@article{Boguslawski2015,
    title={Orbital entanglement in quantum chemistry},
    author={Katharina Boguslawski and Paweł Tecmer},
    journal={International Journal of Quantum Chemistry},
    volume={115},
    number={19},
    pages={1289-1295},
    year={2015},
    doi={10.1002/qua.24832}
}

@article{Williams-Young2023,
    title={A parallel, distributed memory implementation of the adaptive sampling configuration interaction method},
    author={David B. Williams-Young and Norm M. Tubman and Carlos Mejuto-Zaera and Wibe A. de Jong},
    journal={The Journal of Chemical Physics},
    volume={158},
    pages={214109},
    year={2023},
    doi={10.1063/5.0148650},
}

@article{Malvetti2021,
    title={Quantum Circuits for Sparse Isometries},
    author={Emanuel Malvetti and Raban Iten and Roger Colbeck},
    journal={Quantum},
    volume={5},
    pages={412},
    year={2021},
    doi={10.22331/q-2021-06-21-412},
}

@article{Lehtola2018,
    title={Recent developments in libxc — A comprehensive library of functionals for density functional theory},
    author={Susi Lehtola and Conrad Steigemann and Micael J. T. Oliveira and Miguel A. L. Marques},
    journal={SoftwareX},
    volume={7},
    pages={1-5},
    year={2018},
    issn={2352-7110},
    doi={10.1016/j.softx.2017.11.002}
}

@article{Petrone2018,
    title={An Efficient Implementation of Two-Component Relativistic Density Functional Theory with Torque-Free Auxiliary Variables},
    author={Alessio Petrone and David B. Williams--Young and Shichao Sun and Torin F. Stetina and Xiaosong Li},
    journal={The European Physical Journal B},
    volume={91},
    number={169},
    pages={169},
    year={2018},
    doi={10.1140/epjb/e2018-90170-1}
}

@article{Jordan-Wigner1928,
    title={Über das {Paulische} {\"aquivalenzverbot}},
    author={P. Jordan and E. Wigner},
    journal={Zeitschrift für Physik},
    volume={47},
    pages={631-651},
    year={1928},
    doi={10.1007/BF01331938}
}

@article{Love2012,
    title={The {Bravyi-Kitaev} transformation for quantum computation of electronic structure},
    author={Jacob T. Seeley and Martin J. Richard and Peter J. Love},
    journal={The Journal of Chemical Physics},
    volume={137},
    number={22},
    pages={224109},
    year={2012},
    doi={10.1063/1.4768229}
}

@article{Bravyi-Kitaev2002,
    title={Fermionic Quantum Computation},
    author={Sergey B. Bravyi and Alexei Yu. Kitaev},
    journal={Annals of Physics},
    volume={298},
    number={1},
    pages={210-226},
    year={2002},
    issn={0003-4916},
    doi={10.1006/aphy.2002.6254}
}

@article{Christandl2016,
    title={Quantum circuits for isometries},
    author={Raban Iten and Roger Colbeck and Ivan Kukuljan and Jonathan Home and Matthias Christandl},
    journal={Physical Review A},
    volume={93},
    issue={3},
    pages={032318},
    year={2016},
    publisher={American Physical Society},
    doi={10.1103/PhysRevA.93.032318}
}

@article{Kitaev1995,
    title={Quantum measurements and the {Abelian} Stabilizer Problem},
    author={A. Y. Kitaev},
    journal={arXiv},
    year={1995},
    url={https://arxiv.org/abs/quant-ph/9511026}
}

@article{Dobsicek2007,
    title={Arbitrary accuracy iterative quantum phase estimation algorithm using a single ancillary qubit: A two-qubit benchmark},
    author={M. Dobsicek and G. Johansson and V. Shumeiko and G. Wendin},
    journal={Physical Review A},
    volume={76},
    pages={030306},
    year={2007},
    doi={10.1103/PhysRevA.76.030306}
}

@inbook{Nielsen-Chuang2010-QPE,
    booktitle={Quantum Computation and Quantum Information},
    title={The quantum {Fourier} transform and its applications},
    author={Michael A. Nielsen and Isaac L. Chuang},
    year={2010},
    publisher={Cambridge University Press},
    address={Cambridge, UK},
    isbn={978-0-521-63503-0},
    chapter={5.2},
    doi={10.1017/CBO9780511976667}
}

@article{Javadi-Abhari2024,
    title={Quantum computing with {Qiskit}},
    author={Ali Javadi-Abhari and Matthew Treinish and Kevin Krsulich and Christopher J. Wood and Jake Lishman and Julien Gacon and Simon Martiel and Paul D. Nation and Lev S. Bishop and Andrew W. Cross and Blake R. Johnson and Jay M. Gambetta},
    journal={arXiv},
    year={2024},
    doi={10.48550/arXiv.2405.08810}
}

@article{williams2023distributed,
    title={Distributed memory, {GPU} accelerated {Fock} construction for hybrid, {Gaussian} basis density functional theory},
    author={David B. Williams--Young and Andrey Asadchev and Doru Thom Popovici and David Clark and Jonathan Waldrop and Theresa L. Windus and Edward F. Valeev and Wibe A. {de Jong}},
    journal={The Journal of Chemical Physics},
    volume={158},
    number={23},
    pages={234104},
    year={2023},
    doi={10.1063/5.0151070}
}

@article{williams2021achieving,
    title={Achieving performance portability in {Gaussian} basis set density functional theory on accelerator based architectures in {NWChemEx}},
    author={David B. Williams--Young and Abhishek Bagusetty and Wibe A. {de Jong} and Douglas Doerfler and Hubertus J. J. {van Dam} and {\'A}lvaro V{\'a}zquez-Mayagoitia and Theresa L. Windus and Chao Yang},
    journal={Parallel Computing},
    volume={108},
    pages={102829},
    year={2021},
    doi={10.1016/j.parco.2021.102829}
}

@article{williams20on,
    title={On the Efficient Evaluation of the Exchange Correlation Potential on Graphics Processing Unit Clusters},
    author={David B. Williams--Young and Wibe A. {de Jong} and Hubertus J. J. {van Dam} and Chao Yang},
    journal={Frontiers in Chemistry},
    volume={8},
    pages={581058},
    year={2020},
    doi={10.3389/fchem.2020.581058},
}

@misc{Libint2,
    title={Libint: A library for the evaluation of molecular integrals of many-body operators over {Gaussian} functions},
    author={E. F. Valeev},
    howpublished={http://libint.valeyev.net/},
    year={2025}
}

@article{Stein2019,
    title={{autoCAS}: A Program for Fully Automated Multiconfigurational Calculations},
    author={Christopher J. Stein and Markus Reiher},
    journal={Journal of Computational Chemistry},
    volume={40},
    number={25},
    pages={2216-2226},
    year={2019},
    doi={10.1002/jcc.25869},
}

@article{VanVoorhis2002,
    author = {Troy {V}an {V}oorhis and Martin Head-Gordon},
    title = {A geometric approach to direct minimization},
    journal = {Molecular Physics},
    volume = {100},
    number = {11},
    pages = {1713--1721},
    year = {2002},
    doi = {10.1080/00268970110103642},
}

@Inbook{Schlegel1991,
    author={Schlegel, H. B. and McDouall, J. J. W.},
    title={Do You Have {SCF} Stability and Convergence Problems?},
    bookTitle={Computational Advances in Organic Chemistry: Molecular Structure and Reactivity},
    year={1991},
    publisher={Springer Netherlands},
    address={Dordrecht},
    pages={167--185},
    isbn={978-94-011-3262-6},
    doi={10.1007/978-94-011-3262-6_2},
}

@article{Lowdin1956,
  title = {Natural Orbitals in the Quantum Theory of Two-Electron Systems},
  author = {L\"owdin, Per-Olov and Shull, Harrison},
  journal = {Physical Review},
  volume = {101},
  issue = {6},
  pages = {1730--1739},
  numpages = {0},
  year = {1956},
  month = {Mar},
  publisher = {American Physical Society},
  doi = {10.1103/PhysRev.101.1730},
}

@article{Pipek1989,
    author = {Pipek, János and Mezey, Paul G.},
    title = {A fast intrinsic localization procedure applicable for ab initio and semiempirical linear combination of atomic orbital wave functions},
    journal = {The Journal of Chemical Physics},
    volume = {90},
    number = {9},
    pages = {4916-4926},
    year = {1989},
    month = {05},
    issn = {0021-9606},
    doi = {10.1063/1.456588},
}

@article{Edmiston1963,
  title = {Localized Atomic and Molecular Orbitals},
  author = {Edmiston, Clyde and Ruedenberg, Klaus},
  journal = {Rev. Mod. Phys.},
  volume = {35},
  issue = {3},
  pages = {457--464},
  numpages = {0},
  year = {1963},
  month = {Jul},
  publisher = {American Physical Society},
  doi = {10.1103/RevModPhys.35.457},
}

@article{Foster1960,
  title = {Canonical Configurational Interaction Procedure},
  author = {Foster, J. M. and Boys, S. F.},
  journal = {Rev. Mod. Phys.},
  volume = {32},
  issue = {2},
  pages = {300--302},
  numpages = {0},
  year = {1960},
  month = {Apr},
  publisher = {American Physical Society},
  doi = {10.1103/RevModPhys.32.300},
}

@article{Lehtola2013,
    author = {Lehtola, Susi and Jónsson, Hannes},
    title = {Unitary Optimization of Localized Molecular Orbitals},
    journal = {Journal of Chemical Theory and Computation},
    volume = {9},
    number = {12},
    pages = {5365-5372},
    year = {2013},
    doi = {10.1021/ct400793q},
}

@article{Subotnik2005,
    author = {Subotnik, Joseph E. and Dutoi, Anthony D. and Head-Gordon, Martin},
    title = {Fast localized orthonormal virtual orbitals which depend smoothly on nuclear coordinates},
    journal = {The Journal of Chemical Physics},
    volume = {123},
    number = {11},
    pages = {114108},
    year = {2005},
    month = {09},
    issn = {0021-9606},
    doi = {10.1063/1.2033687},
}

@article{Wang2025,
    author = {Wang, Zhenling and Ikeda, Kevin and Shen, Hengyuan and Loipersberger, Matthias and Zech, Alexander and Aldossary, Abdulrahman and Head-Gordon, Teresa and Head-Gordon, Martin},
    title = {Second-Generation Energy Decomposition Analysis of Intermolecular Interaction Energies from the Second-Order {M{\o}ller--Plesset} Theory: An Extensible, Orthogonal Formulation with Useful Basis Set Convergence for All Terms},
    journal = {Journal of Chemical Theory and Computation},
    volume = {21},
    number = {3},
    pages = {1163-1178},
    year = {2025},
    doi = {10.1021/acs.jctc.4c01301},
}

@article{Tubman2016,
    author = {Tubman, Norm M. and Lee, Joonho and Takeshita, Tyler Y. and Head-Gordon, Martin and Whaley, K. Birgitta},
    title = {A deterministic alternative to the full configuration interaction quantum {Monte Carlo} method},
    journal = {The Journal of Chemical Physics},
    volume = {145},
    number = {4},
    pages = {044112},
    year = {2016},
    month = {07},
    doi = {10.1063/1.4955109},
}

@article{Tubman2020,
    author = {Tubman, Norm M. and Freeman, C. Daniel and Levine, Daniel S. and Hait, Diptarka and Head-Gordon, Martin and Whaley, K. Birgitta},
    title = {Modern Approaches to Exact Diagonalization and Selected Configuration Interaction with the Adaptive Sampling {CI} Method},
    journal = {Journal of Chemical Theory and Computation},
    volume = {16},
    number = {4},
    pages = {2139-2159},
    year = {2020},
    doi = {10.1021/acs.jctc.8b00536},
}

@article{Shaw2017,
    author = {Shaw, Robert A. and Hill, J. Grant},
    doi = {10.1063/1.4986887},
    issn = {0021-9606},
    journal = {The Journal of Chemical Physics},
    month = {aug},
    number = {7},
    pages = {074108},
    title = {Prescreening and efficiency in the evaluation of integrals over ab initio effective core potentials},
    volume = {147},
    year = {2017}
}

@article{Shaw2021,
    author = {Robert A. Shaw and J. Grant Hill},
    doi = {10.21105/joss.03039},
    journal = {Journal of Open Source Software},
    publisher = {The Open Journal},
    number = {60},
    pages = {3039},
    title = {libecpint: A {C++} library for the efficient evaluation of integrals over effective core potentials},
    volume = {6},
    year = {2021}
}

@article{Sun2020,
    author = {Sun, Qiming and Zhang, Xing and Banerjee, Samragni and Bao, Peng and Barbry, Marc and Blunt, Nick S. and Bogdanov, Nikolay A. and Booth, George H. and Chen, Jia and Cui, Zhi-Hao and Eriksen, Janus J. and Gao, Yang and Guo, Sheng and Hermann, Jan and Hermes, Matthew R. and Koh, Kevin and Koval, Peter and Lehtola, Susi and Li, Zhendong and Liu, Junzi and Mardirossian, Narbe and McClain, James D. and Motta, Mario and Mussard, Bastien and Pham, Hung Q. and Pulkin, Artem and Purwanto, Wirawan and Robinson, Paul J. and Ronca, Enrico and Sayfutyarova, Elvira R. and Scheurer, Maximilian and Schurkus, Henry F. and Smith, James E. T. and Sun, Chong and Sun, Shi-Ning and Upadhyay, Shiv and Wagner, Lucas K. and Wang, Xiao and White, Alec and Whitfield, James Daniel and Williamson, Mark J. and Wouters, Sebastian and Yang, Jun and Yu, Jason M. and Zhu, Tianyu and Berkelbach, Timothy C. and Sharma, Sandeep and Sokolov, Alexander Yu. and Chan, Garnet Kin-Lic},
    title = {Recent developments in the {PySCF} program package},
    journal = {The Journal of Chemical Physics},
    volume = {153},
    number = {2},
    pages = {024109},
    year = {2020},
    month = {07},
    issn = {0021-9606},
    doi = {10.1063/5.0006074},
}

@article{Lehtola2023,
    author = {Lehtola, Susi},
    title = {A call to arms: Making the case for more reusable libraries},
    journal = {The Journal of Chemical Physics},
    volume = {159},
    number = {18},
    pages = {180901},
    year = {2023},
    month = {11},
    issn = {0021-9606},
    doi = {10.1063/5.0175165},
}

@article{Smith2021,
    author = {Smith, Daniel G. A. and Altarawy, Doaa and Burns, Lori A. and Welborn, Matthew and Naden, Levi N. and Ward, Logan and Ellis, Sam and Pritchard, Benjamin P. and Crawford, T. Daniel},
    title = {The {MolSSI} {QCArchive} project: An open-source platform to compute, organize, and share quantum chemistry data},
    journal = {WIREs Computational Molecular Science},
    volume = {11},
    number = {2},
    pages = {e1491},
    doi = {10.1002/wcms.1491},
    year = {2021}
}

@article{Weymuth2024,
    author = {Weymuth, Thomas and Unsleber, Jan P. and Türtscher, Paul L. and Steiner, Miguel and Sobez, Jan-Grimo and Müller, Charlotte H. and Mörchen, Maximilian and Klasovita, Veronika and Grimmel, Stephanie A. and Eckhoff, Marco and Csizi, Katja-Sophia and Bosia, Francesco and Bensberg, Moritz and Reiher, Markus},
    title = {{SCINE}—Software for chemical interaction networks},
    journal = {The Journal of Chemical Physics},
    volume = {160},
    number = {22},
    pages = {222501},
    year = {2024},
    month = {06},
    issn = {0021-9606},
    doi = {10.1063/5.0206974},
}

@article{Pizzi2016,
  author       = {Giovanni Pizzi and Andrea Cepellotti and Riccardo Sabatini
                  and Nicola Marzari and Boris Kozinsky},
  title        = {{AiiDA}: automated interactive infrastructure and database
                  for computational science},
  journal      = {Computational Materials Science},
  volume       = {111},
  pages        = {218--230},
  year         = {2016},
  doi          = {10.1016/j.commatsci.2015.09.013},
}

@article{Huber2020,
  author       = {Sebastiaan P. Huber and Spyros Zoupanos and Martin Uhrin
                  and Leopold Talirz and Leonid Kahle and Rico Häuselmann
                  and Dominik Gresch and Tiziano Müller and Aliaksandr V. Yakutovich
                  and Casper W. Andersen and Francisco F. Ramirez and Carl S. Adorf
                  and Fernando Gargiulo and Snehal Kumbhar and Elsa Passaro
                  and Conrad Johnston and Andrius Merkys and Andrea Cepellotti
                  and Nicolas Mounet and Nicola Marzari and Boris Kozinsky
                  and Giovanni Pizzi},
  title        = {{AiiDA 1.0}, a scalable computational infrastructure for automated
                  reproducible workflows and data provenance},
  journal      = {Scientific Data},
  volume       = {7},
  page         = {300},
  year         = {2020},
  doi          = {10.1038/s41597-020-00638-4},
}

@article{Uhrin2021,
  author       = {Martin Uhrin and Sebastiaan P. Huber and Jusong Yu
                  and Nicola Marzari and Giovanni Pizzi},
  title        = {Workflows in {AiiDA}: Engineering a high-throughput,
                  event-based engine for robust and modular computational workflows},
  journal      = {Computational Materials Science},
  volume       = {187},
  page         = {110086},
  year         = {2021},
  doi          = {10.1016/j.commatsci.2020.110086},
}

@article{Scheidgen2023,
  author       = {Scheidgen, Christian and others},
  title        = {{NOMAD}: A distributed web-based platform
                  for managing materials science research data},
  journal      = {Journal of Open Source Software},
  volume       = {8},
  number       = {90},
  year         = {2023},
  doi          = {10.21105/joss.05388},
}

@article{Alvarez2015,
  author  = {\'{A}lvarez‑Moreno, M. and de Graaf, C. and L\'{o}pez, N. and Maseras, F. and Poblet, J. M. and Bo, C.},
  title   = {Managing the Computational Chemistry Big Data Problem: The {ioChem-BD} Platform},
  journal = {Journal of Chemical Information and Modeling},
  year    = {2015},
  volume  = {55},
  number  = {1},
  pages   = {95-103},
  doi     = {10.1021/ci500593j},
}

@misc{Wenzel2025,
  author       = {Wenzel Jakob and
                  Henry Schreiner and
                  Jason Rhinelander and
                  Ralf W. Grosse-Kunstleve and
                  Dean Moldovan and
                  Ivan Smirnov and
                  Aaron Gokaslan and
                  Yannick Jadoul and
                  Axel Huebl and
                  Boris Staletic and
                  Sergei Izmailov and
                  Eric Cousineau and
                  Dustin Spicuzza and
                  Michael Carlstrom and
                  Bruce Merry and
                  b-pass and
                  Antony Lee and
                  Sylvain Corlay and
                  Lori A. Burns and
                  Dan and
                  Xuehai Pan and
                  bennorth and
                  Trent Houliston and
                  Sergey Lyskov and
                  Robert Haschke and
                  jbarlow and
                  gentlegiantJGC and
                  Michael Šimáček},
  title        = {pybind/pybind11: Version 3.0.1},
  year         = {2025},
  publisher    = {Zenodo},
  version      = {v3.0.1},
  doi          = {10.5281/zenodo.16929811},
}

@article{Larsen2017,
  author       = {Larsen, Ask Hjorth and Mortensen, Jens J{\o}rgen and Blomqvist, Jakob and Castelli, Ivano E. and Christensen, Rune and Du{\l}ak, Marcin and Friis, Jesper and Groves, Michael N. and Hammer, Bj{\o}rk and Hargus, Cory and Hermes, Eric D. and Jennings, Paul C. and Jensen, Peter B. and Kermode, James and Kitchin, John R. and Kolsbjerg, Esben Leonhard and Kubal, Joseph and Kaasbjerg, Kristen and Lysgaard, Steen and Maronsson, J{\'o}n Bergmann and Maxson, Tristan and Olsen, Thomas and Pastewka, Lars and Peterson, Andrew and Rostgaard, Carsten and Schi{\o}tz, Jakob and Schütt, Ole and Strange, Mikkel and Thygesen, Kristian S. and Vegge, Tejs and Vilhelmsen, Lasse and Walter, Michael and Zeng, Zhenhua and Jacobsen, Karsten W.},
  title        = {The {Atomic Simulation Environment}—a {Python} library for working with atoms},
  journal      = {Journal of Physics: Condensed Matter},
  volume       = {29},
  number       = {27},
  pages        = {273002},
  year         = {2017},
  doi          = {10.1088/1361-648X/aa680e},
}

@incollection{Hatano2005,
    title={Finding exponential product formulas of higher orders},
    author={Hatano, Naomichi and Suzuki, Masuo},
    booktitle={Quantum annealing and other optimization methods},
    pages={37--68},
    year={2005},
    publisher={Springer}
}

@article{Dutkiewicz2025,
    title = {Error Mitigation and Circuit Division for Early Fault-Tolerant Quantum Phase Estimation},
    author = {Dutkiewicz, Alicja and Polla, Stefano and Scheurer, Maximilian and Gogolin, Christian and Huggins, William J. and {O'Brien}, Thomas E.},
    journal = {PRX Quantum},
    volume = {6},
    issue = {4},
    pages = {040318},
    numpages = {36},
    year = {2025},
    month = {Oct},
    publisher = {American Physical Society},
    doi = {10.1103/mlmy-yskj},
}

@misc{QDK,
   author = {{Microsoft}},
   year = {2017},
   title = {{Azure Quantum Development Kit}},
   howpublished = {\url{https://github.com/microsoft/qsharp}}
}

@misc{VSCode,
   author = {{Microsoft}},
   title = {{Visual Studio Code}},
   howpublished = {\url{https://code.visualstudio.com/}}
}

@article{Richard2023,
    author = {Richard, Ryan M. and Keipert, Kristopher and Waldrop, Jonathan and Keçeli, Murat and Williams-Young, David and Bair, Raymond and Boschen, Jeffery and Crandall, Zachery and Gasperich, Kevin and Mahmud, Quazi Ishtiaque and Panyala, Ajay and Valeev, Edward and van Dam, Hubertus and de Jong, Wibe A. and Windus, Theresa L.},
    title = {{PluginPlay}: Enabling exascale scientific software one module at a time},
    journal = {The Journal of Chemical Physics},
    volume = {158},
    number = {18},
    pages = {184801},
    year = {2023},
    month = {05},
    issn = {0021-9606},
    doi = {10.1063/5.0147903},
}

@article{Blum2024,
	title = {Roadmap on methods and software for electronic structure based simulations in chemistry and materials},
	volume = {6},
	issn = {2516-1075},
	doi = {10.1088/2516-1075/ad48ec},
	number = {4},
	urldate = {2026-01-11},
	journal = {Electronic Structure},
	author = {Blum, Volker and Asahi, Ryoji and Autschbach, Jochen and Bannwarth, Christoph and Bihlmayer, Gustav and Blügel, Stefan and Burns, Lori A and Crawford, T Daniel and Dawson, William and De Jong, Wibe Albert and Draxl, Claudia and Filippi, Claudia and Genovese, Luigi and Giannozzi, Paolo and Govind, Niranjan and Hammes-Schiffer, Sharon and Hammond, Jeff R and Hourahine, Benjamin and Jain, Anubhav and Kanai, Yosuke and Kent, Paul R C and Larsen, Ask Hjorth and Lehtola, Susi and Li, Xiaosong and Lindh, Roland and Maeda, Satoshi and Makri, Nancy and Moussa, Jonathan and Nakajima, Takahito and Nash, Jessica A and Oliveira, Micael J T and Patel, Pansy D and Pizzi, Giovanni and Pourtois, Geoffrey and Pritchard, Benjamin P and Rabani, Eran and Reiher, Markus and Reining, Lucia and Ren, Xinguo and Rossi, Mariana and Schlegel, H Bernhard and Seriani, Nicola and Slipchenko, Lyudmila V and Thom, Alexander and Valeev, Edward F and Van Troeye, Benoit and Visscher, Lucas and Vlček, Vojtěch and Werner, Hans-Joachim and Williams-Young, David B and Windus, Theresa L.},
	month = dec,
	year = {2024},
	pages = {042501},
}

@article{DiFelice2023,
author = {Di Felice, Rosa and Mayes, Maricris L. and Richard, Ryan M. and Williams-Young, David B. and Chan, Garnet Kin-Lic and de Jong, Wibe A. and Govind, Niranjan and Head-Gordon, Martin and Hermes, Matthew R. and Kowalski, Karol and Li, Xiaosong and Lischka, Hans and Mueller, Karl T. and Mutlu, Erdal and Niklasson, Anders M. N. and Pederson, Mark R. and Peng, Bo and Shepard, Ron and Valeev, Edward F. and van Schilfgaarde, Mark and Vlaisavljevich, Bess and Windus, Theresa L. and Xantheas, Sotiris S. and Zhang, Xing and Zimmerman, Paul M.},
title = {A Perspective on Sustainable Computational Chemistry Software Development and Integration},
journal = {Journal of Chemical Theory and Computation},
volume = {19},
number = {20},
pages = {7056-7076},
year = {2023},
doi = {10.1021/acs.jctc.3c00419},
}

@article{Kowalski2021,
author = {Kowalski, Karol and Bair, Raymond and Bauman, Nicholas P. and Boschen, Jeffery S. and Bylaska, Eric J. and Daily, Jeff and de Jong, Wibe A. and Dunning, Thom Jr. and Govind, Niranjan and Harrison, Robert J. and Ke{\c{c}}eli, Murat and Keipert, Kristopher and Krishnamoorthy, Sriram and Kumar, Suraj and Mutlu, Erdal and Palmer, Bruce and Panyala, Ajay and Peng, Bo and Richard, Ryan M. and Straatsma, T. P. and Sushko, Peter and Valeev, Edward F. and Valiev, Marat and van Dam, Hubertus J. J. and Waldrop, Jonathan M. and Williams-Young, David B. and Yang, Chao and Zalewski, Marcin and Windus, Theresa L.},
title = {From {NWChem} to {NWChemEx}: Evolving with the Computational Chemistry Landscape},
journal = {Chemical Reviews},
volume = {121},
number = {8},
pages = {4962-4998},
year = {2021},
doi = {10.1021/acs.chemrev.0c00998},
}

@article{Berquist2024,
    author = {Berquist, Eric and Dumi, Amanda and Upadhyay, Shiv and Abarbanel, Omri D. and Cho, Minsik and Gaur, Sagar and Gil, Renee and Hutchison, Geoffrey R. and Lee, Oliver S. and Rosen, Andrew S. and Schamnad, Sanjeed and Schneider, Felipe S. S. and Steinmann, Casper and Stolyarchuk, Maxim and Vandezande, Jonathon E. and Zak, Weronika and Langner, Karol M.},
    title = {{cclib} 2.0: An updated architecture for
                    interoperable computational chemistry},
    journal = {The Journal of Chemical Physics},
    volume = {161},
    number = {4},
    pages = {042501},
    year = {2024},
    month = {07},
    issn = {0021-9606},
    doi = {10.1063/5.0216778},
}

@article{Barnes2024,
    author = {Barnes, T. A. and Ellis, S. and Chen, J. and Plimpton, S. J. and Nash, J. A.},
    title = {Plugin-based interoperability and ecosystem management for the {MolSSI} Driver Interface Project},
    journal = {The Journal of Chemical Physics},
    volume = {160},
    number = {21},
    pages = {214114},
    year = {2024},
    month = {06},
    issn = {0021-9606},
    doi = {10.1063/5.0214279},
}

@article{McClean2020,
    author = {McClean, Jarrod R. and Rubin, Nicholas C. and Sung, Kevin J. and Kivlichan, Ian D. and Bonet-Monroig, Xavier and Cao, Yudong and Dai, Chengyu and Fried, E. Schuyler and Gidney, Craig and Gimber, Brendan and Gokhale, Pranav and Häner, Thomas and Harber, Tarini and Harrigan, Matthew P. and Hastad, Jennifer and Higgott, Oscar and Jiang, Zhang and Khatri, Sumeet and Kieferová, Mária and Koh, Dax Enshan and Kothari, Robin and Li, Guang Hao Low and Liu, Yu-An and McArdle, Sam and Motta, Mario and {O'Brien}, Thomas E. and Peropadre, Borja and Rubin, Nicholas C. and Sawaya, Nicolas P. D. and Setia, Kanav and Sim, Sukin and Steiger, Damian S. and Tubman, Norm M. and Vanapalli, Sarada and Wecker, David and Wiebe, Nathan and Yamamoto, Takeshi and Yao, Jiajun and Zhang, Fang and Babbush, Ryan},
    title = {{OpenFermion}: the electronic structure package for quantum computers},
    journal = {Quantum Science and Technology},
    volume = {5},
    number = {3},
    pages = {034014},
    year = {2020},
    doi = {10.1088/2058-9565/ab8ebc},
}

@article{Smith2020Psi4,
    author = {Smith, Daniel G. A. and Burns, Lori A. and Simmonett, Andrew C. and Parrish, Robert M. and Schieber, Matthew C. and Galvelis, Raimondas and Kraus, Peter and Kruse, Holger and Di Remigio, Roberto and Alenaizan, Asem and James, Andrew M. and Lehtola, Susi and Misiewicz, Jonathon P. and Scheurer, Maximilian and Shaw, Robert A. and Schriber, Jeffrey B. and Xie, Yi and Glick, Zachary L. and Sirianni, Dominic A. and {O'Brien}, Joseph Senan and Waldrop, Jonathan M. and Kumar, Ashutosh and Hohenstein, Edward G. and Pritchard, Benjamin P. and Brooks, Bernard R. and Schaefer III, Henry F. and Sokolov, Alexander Yu. and Patkowski, Konrad and DePrince III, A. Eugene and Bozkaya, Uğur and King, Rollin A. and Evangelista, Francesco A. and Turney, Justin M. and Crawford, T. Daniel and Sherrill, C. David},
    title = {Psi4 1.4: Open-source software for high-throughput quantum chemistry},
    journal = {The Journal of Chemical Physics},
    volume = {152},
    number = {18},
    pages = {184108},
    year = {2020},
    doi = {10.1063/5.0006002},
}

@article{Kottmann2021,
    author = {Kottmann, Jakob S. and Alperin-Lea, Sumner and Tamayo-Mendoza, Teresa and Cervera-Lierta, Alba and Lavigne, Cyrille and Yen, Tzu-Ching and Verteletskyi, Vladyslav and Schleich, Philipp and Anand, Abhinav and Degroote, Matthias and Chaez, Skylar and Grimsley, Harper R. and Mayer, Nicholas and Muecke, Lennart and Izmaylov, Artur F.},
    title = {{Tequila}: a platform for rapid development of quantum algorithms},
    journal = {Quantum Science and Technology},
    volume = {6},
    number = {2},
    pages = {024009},
    year = {2021},
    doi = {10.1088/2058-9565/abe567},
}

@article{Senicourt2022,
    author = {Senicourt, Valentin and Brown, James and Fleury, Alexandre and Krauss, Ryan and Gao, Yukun and Rubin, Nicholas and Benchen, Artur and Dinh, Hung and Liepuoniute, Ieva and Parrish, Robert and Gao, Xuelan and Rice, Julia},
    title = {{Tangelo}: An Open-source Python Package for End-to-end Chemistry Workflows on Quantum Computers},
    journal = {arXiv},
    year = {2022},
    url = {https://arxiv.org/abs/2206.12424},
}

@article{McArdle2020,
    author = {McArdle, Sam and Endo, Suguru and Aspuru-Guzik, Al\'an and Benjamin, Simon C. and Yuan, Xiao},
    title = {Quantum computational chemistry},
    journal = {Reviews of Modern Physics},
    volume = {92},
    issue = {1},
    pages = {015003},
    year = {2020},
    doi = {10.1103/RevModPhys.92.015003},
}

@article{McCaskey2020,
    author = {McCaskey, Alexander J. and Lyakh, Dmitry I. and Dumitrescu, Eugene F. and Powers, Sarah S. and Humble, Travis S.},
    title = {{XACC}: a system-level software infrastructure for heterogeneous quantum-classical computing},
    journal = {Quantum Science and Technology},
    volume = {5},
    number = {2},
    pages = {024002},
    year = {2020},
    doi = {10.1088/2058-9565/ab6bf6},
}

@book{Dalzell2025,
    author = {Dalzell, Alexander M. and McArdle, Sam and Berta, Mario and Bienias, Przemyslaw and Chen, Chi-Fang and Gilyén, András and Hann, Connor T. and Kastoryano, Michael J. and Khabiboulline, Emil T. and Kubica, Aleksander and Salton, Grant and Wang, Samson and Brandão, Fernando G. S. L.},
    title = {Quantum Algorithms: A Survey of Applications and End-to-End Complexities},
    publisher = {Cambridge University Press},
    year = {2025},
    doi = {10.1017/9781009490061},
}

@article{Bergholm2022,
    author = {Bergholm, Ville and Izaac, Josh and Schuld, Maria and Gogolin, Christian and Ahmed, Shahnawaz and Ajber, Vishnu and Sohaib Alam, Muhammad and Alonso-Linaje, Guillermo and AkashNarayanan, B. and Asadi, Ali and Arrazola, Juan Miguel and Azad, Utkarsh and Banning, Samuel and Blank, Carsten and Bromley, Thomas R. and Cordier, Benjamin A. and Ceroni, Jack and Delgado, Alain and Di Matteo, Olivia and Duber, Amintor and Dwiputra, Anurag and Ekholm, Elina and Endicott, Philipp and Feynman, Richard P. and Fiorentini, Stefano and Fullwood, James A. and Gadipudi, Karun and Giordano, Nicolò and Goldberg, Noah and Guala, Dharmik and Gyurik, Casper and Huet, Louis and Isakov, Mikhail and Izquierdo, Víctor and Jahangiri, Soran and Killoran, Nathan and Kumar, Ashish and Zaldívar Lami, Cody and Lazzarin, Maria and Lee, Christina and Li, Elica and Lopatnikov, Anton and Lowe, Angus and McAdams, Kelly and McCormick, James and Mueller, Timo E. and O'Riordan, Lee James and Pacher, Carsten and Pardini, Lana and Park, S. Kavitha and Patel, Ishamel and Paulos, Aaron and Pavlović, Borja and Pelofske, Elijah and Pesah, Arthur and Pietrucci, Romain and Pinzón, David Guillermo and Ponturo, Alberto and Quesada, Nicolás and Roberts, Chase and Rodriguez, Israel and Rose, Antal and Rowe, Nate and Sabo, Filippo and Schuld, Maria and Schwindt, Peter and Shankar, Jayant and Stanwyck, Samuel and Steckmann, Thaddäus and Sud, Ejaaz and Tamayo-Mendoza, Teresa and Tan, Wei and Tao, Yue and Thompson, Glenn and Torres, Alberto and Trotzky, Stefan and Vargas-Hernández, Rodrigo A. and Vidal, Javier and Vincent, Trevor and Vujosevic, Lazar and Vidal, Javier and Wang, David and Wiersema, Roeland and Wipf, Martin and Woźniak, Piotr Jerzy and Zhang, Haodong and Zhu, Shaoming},
    title = {PennyLane: Automatic differentiation of hybrid quantum-classical computations},
    journal = {arXiv},
    year = {2022},
    doi = {10.48550/arXiv.1811.04968}
}

@misc{Cirq2024,
    author = {{Cirq Developers}},
    title = {Cirq},
    year = {2024},
    doi = {10.5281/zenodo.10247207},
    howpublished = {\url{https://github.com/quantumlib/Cirq}},
}

@article{Zhang2025TrimCI,
    author = {Zhang, Hao and Otten, Matthew},
    title = {From Random Determinants to the Ground State},
    journal = {arXiv},
    year = {2025},
    eprint = {2511.14734},
    archivePrefix = {arXiv},
    primaryClass = {quant-ph},
    doi = {10.48550/arXiv.2511.14734},
}

@article{Childs2012,
  author = {Andrew M. Childs and Nathan Wiebe},
  title = {Hamiltonian Simulation Using Linear Combinations of Unitary Operations},
  journal = {Quantum Information and Computation},
  volume = {12},
  number = {11-12},
  pages = {901--924},
  year = {2012},
  eprint = {1202.5822},
  archivePrefix = {arXiv},
  primaryClass = {quant-ph},
}

@article{Chakraborty2024,
  author = {Shantanav Chakraborty},
  title = {Implementing any Linear Combination of Unitaries on Intermediate-term Quantum Computers},
  journal = {Quantum},
  volume = {8},
  pages = {1496},
  year = {2024},
  doi = {10.22331/q-2024-10-10-1496},
  eprint = {2302.13555},
  archivePrefix = {arXiv},
  primaryClass = {quant-ph},
}

@article{Sanders2019,
  author = {Yuval R. Sanders and Guang Hao Low and Artur Scherer and Dominic W. Berry},
  title = {Black-Box Quantum State Preparation without Arithmetic},
  journal = {Physical Review Letters},
  volume = {122},
  pages = {020502},
  year = {2019},
  doi = {10.1103/PhysRevLett.122.020502},
  eprint = {1807.03206},
  archivePrefix = {arXiv},
  primaryClass = {quant-ph},
}

@article{Fomichev2024,
  author = {Stepan Fomichev and Kasra Hejazi and Modjtaba Shokrian Zini and Matthew Kiser and Joana Fraxanet and Pablo Antonio Moreno Casares and Alain Delgado and Joonsuk Huh and Arne-Christian Voigt and Jonathan E. Mueller and Juan Miguel Arrazola},
  title = {Initial State Preparation for Quantum Chemistry on Quantum Computers},
  journal = {PRX Quantum},
  volume = {5},
  number = {4},
  pages = {040339},
  year = {2024},
  month = {Dec},
  doi = {10.1103/PRXQuantum.5.040339},
  eprint = {2310.18410},
  archivePrefix = {arXiv},
  primaryClass = {quant-ph},
}

@article{Pulay1988,
  author = {Pulay, Peter and Hamilton, Tracy P.},
  title = {{UHF} natural orbitals for defining and starting {MC-SCF} calculations},
  journal = {The Journal of Chemical Physics},
  volume = {88},
  number = {7},
  pages = {4926--4933},
  year = {1988},
  doi = {10.1063/1.454704}
}

@article{Khedkar2019,
  author = {Khedkar, Abhishek and Roemelt, Michael},
  title = {Active Space Selection Based on Natural Orbital Occupation Numbers from n-Electron Valence Perturbation Theory},
  journal = {Journal of Chemical Theory and Computation},
  volume = {15},
  number = {6},
  pages = {3522--3536},
  year = {2019},
  doi = {10.1021/acs.jctc.8b01293}
}

@INPROCEEDINGS{Liu2025,
  author={Liu, Yuhao and Yao, Kevin and Hong, Jonathan and Froustey, Julien and Rrapaj, Ermal and Iancull, Costin and Li, Gushu and Shi, Yunong},
  booktitle={2025 IEEE International Symposium on High Performance Computer Architecture (HPCA)}, 
  title={HATT: Hamiltonian Adaptive Ternary Tree for Optimizing Fermion-to-Qubit Mapping}, 
  year={2025},
  volume={},
  number={},
  pages={143-157},
  doi={10.1109/HPCA61900.2025.00022}}

@article{Hoefler2023,
	author = {Hoefler, Torsten and Häner, Thomas and Troyer, Matthias},
	title = {Disentangling Hype from Practicality: On Realistically Achieving Quantum Advantage},
	volume = {66},
	doi = {10.1145/3571725},
	number = {5},
	journal = {Communications of the ACM},
	month = may,
	year = {2023},
	pages = {82--87},
}

@article{Huber2021,
	title = {Common workflows for computing material properties using different quantum engines},
	volume = {7},
	issn = {2057-3960},
	doi = {10.1038/s41524-021-00594-6},
	number = {1},
	urldate = {2026-01-12},
	journal = {npj Computational Materials},
	author = {Huber, Sebastiaan P. and Bosoni, Emanuele and Bercx, Marnik and Bröder, Jens and Degomme, Augustin and Dikan, Vladimir and Eimre, Kristjan and Flage-Larsen, Espen and Garcia, Alberto and Genovese, Luigi and Gresch, Dominik and Johnston, Conrad and Petretto, Guido and Poncé, Samuel and Rignanese, Gian-Marco and Sewell, Christopher J. and Smit, Berend and Tseplyaev, Vasily and Uhrin, Martin and Wortmann, Daniel and Yakutovich, Aliaksandr V. and Zadoks, Austin and Zarabadi-Poor, Pezhman and Zhu, Bonan and Marzari, Nicola and Pizzi, Giovanni},
	month = aug,
	year = {2021},
	pages = {136},
}

@article{Aspuru-Guzik2005,
	title = {Simulated Quantum Computation of Molecular Energies},
	volume = {309},
	issn = {0036-8075, 1095-9203},
	doi = {10.1126/science.1113479},
	number = {5741},
	urldate = {2026-01-12},
	journal = {Science},
	author = {Aspuru-Guzik, Alán and Dutoi, Anthony D. and Love, Peter J. and Head-Gordon, Martin},
	month = sep,
	year = {2005},
	pages = {1704--1707},
}

@article{Lanyon2010,
	title = {Towards quantum chemistry on a quantum computer},
	volume = {2},
	issn = {1755-4330, 1755-4349},
	doi = {10.1038/nchem.483},
	number = {2},
	urldate = {2026-01-12},
	journal = {Nature Chemistry},
	author = {Lanyon, B. P. and Whitfield, J. D. and Gillett, G. G. and Goggin, M. E. and Almeida, M. P. and Kassal, I. and Biamonte, J. D. and Mohseni, M. and Powell, B. J. and Barbieri, M. and Aspuru-Guzik, A. and White, A. G.},
	month = feb,
	year = {2010},
	pages = {106--111},
}

@article{Svore2014,
author = {Svore, Krysta M. and Hastings, Matthew B. and Freedman, Michael},
title = {Faster phase estimation},
year = {2014},
issue_date = {March 2014},
publisher = {Rinton Press, Incorporated},
volume = {14},
number = {3--4},
issn = {1533-7146},
journal = {Quantum Info. Comput.},
month = mar,
pages = {306--328},
numpages = {23},
}

@article{Alexeev2025,
	title = {A Perspective on Quantum Computing Applications in Quantum Chemistry Using 25--100 Logical Qubits},
	author = {Alexeev, Yuri and Batista, Victor S. and Bauman, Nicholas and Bertels, Luke and Claudino, Daniel and Dutta, Rishab and Gagliardi, Laura and Godwin, Scott and Govind, Niranjan and Head-Gordon, Martin and Hermes, Matthew R. and Kowalski, Karol and Li, Ang and Liu, Chenxu and Liu, Junyu and Liu, Ping and García-Lastra, Juan M. and Mejia-Rodriguez, Daniel and Mueller, Karl and Otten, Matthew and Peng, Bo and Raugas, Mark and Reiher, Markus and Rigor, Paul and Shaw, Wendy J. and Van Schilfgaarde, Mark and Vegge, Tejs and Zhang, Yu and Zheng, Muqing and Zhu, Linghua},
  volume = {21},
	doi = {10.1021/acs.jctc.5c01038},
	language = {en},
	number = {22},
	urldate = {2026-01-12},
	journal = {Journal of Chemical Theory and Computation},
	month = nov,
	year = {2025},
	pages = {11335--11357},
}

@article{Aasen2025,
	title = {Roadmap to fault tolerant quantum computation using topological qubit arrays},
	copyright = {Creative Commons Attribution 4.0 International},
	doi = {10.48550/arxiv.2502.12252},
	urldate = {2026-01-12},
	publisher = {arXiv},
	author = {Aasen, David and Aghaee, Morteza and Alam, Zulfi and Andrzejczuk, Mariusz and Antipov, Andrey and Astafev, Mikhail and Avilovas, Lukas and Barzegar, Amin and Bauer, Bela and Becker, Jonathan and Bello-Rivas, Juan M. and Bhaskar, Umesh and Bocharov, Alex and Boddapati, Srini and Bohn, David and Bommer, Jouri and Bonderson, Parsa and Borovsky, Jan and Bourdet, Leo and Boutin, Samuel and Brown, Tom and Campbell, Gary and Casparis, Lucas and Chakravarthi, Srivatsa and Chao, Rui and Chapman, Benjamin J. and Chatoor, Sohail and Christensen, Anna Wulff and Codd, Patrick and Cole, William and Cooper, Paul and Corsetti, Fabiano and Cui, Ajuan and van Dam, Wim and Dandachi, Tareq El and Daraeizadeh, Sahar and Dumitrascu, Adrian and Ekefjärd, Andreas and Fallahi, Saeed and Galletti, Luca and Gardner, Geoff and Gatta, Raghu and Gavranovic, Haris and Goulding, Michael and Govender, Deshan and Griggio, Flavio and Grigoryan, Ruben and Grijalva, Sebastian and Gronin, Sergei and Gukelberger, Jan and Haah, Jeongwan and Hamdast, Marzie and Hansen, Esben Bork and Hastings, Matthew and Heedt, Sebastian and Ho, Samantha and Hogaboam, Justin and Holgaard, Laurens and Van Hoogdalem, Kevin and Indrapiromkul, Jinnapat and Ingerslev, Henrik and Ivancevic, Lovro and Jablonski, Sarah and Jensen, Thomas and Jhoja, Jaspreet and Jones, Jeffrey and Kalashnikov, Kostya and Kallaher, Ray and Kalra, Rachpon and Karimi, Farhad and Karzig, Torsten and Kimes, Seth and Kliuchnikov, Vadym and Kloster, Maren Elisabeth and Knapp, Christina and Knee, Derek and Koski, Jonne and Kostamo, Pasi and Kuesel, Jamie and Lackey, Brad and Laeven, Tom and Lai, Jeffrey and de Lange, Gijs and Larsen, Thorvald and Lee, Jason and Lee, Kyunghoon and Leum, Grant and Li, Kongyi and Lindemann, Tyler and Lucas, Marijn and Lutchyn, Roman and Madsen, Morten Hannibal and Madulid, Nash and Manfra, Michael and Markussen, Signe Brynold and Martinez, Esteban and Mattila, Marco and Mattinson, Jake and McNeil, Robert and Mei, Antonio Rodolph and Mishmash, Ryan V. and Mohandas, Gopakumar and Mollgaard, Christian and de Moor, Michiel and Morgan, Trevor and Moussa, George and Narla, Anirudh and Nayak, Chetan and Nielsen, Jens Hedegaard and Nielsen, William Hvidtfelt Padkær and Nolet, Frédéric and Nystrom, Mike and O'Farrell, Eoin and Otani, Keita and Paetznick, Adam and Papon, Camille and Paz, Andres and Petersson, Karl and Petit, Luca and Pikulin, Dima and Pons, Diego Olivier Fernandez and Quinn, Sam and Rajpalke, Mohana and Ramirez, Alejandro Alcaraz and Rasmussen, Katrine and Razmadze, David and Reichardt, Ben and Ren, Yuan and Reneris, Ken and Riccomini, Roy and Sadovskyy, Ivan and Sainiemi, Lauri and Saldaña, Juan Carlos Estrada and Sanlorenzo, Irene and Schaal, Simon and Schmidgall, Emma and Sfiligoj, Cristina and da Silva, Marcus P. and Singh, Shilpi and Sinha, Sarat and Soeken, Mathias and Sohr, Patrick and Stankevic, Tomas and Stek, Lieuwe and Strøm-Hansen, Patrick and Stuppard, Eric and Sundaram, Aarthi and Suominen, Henri and Suter, Judith and Suzuki, Satoshi and Svore, Krysta and Teicher, Sam and Thiyagarajah, Nivetha and Tholapi, Raj and Thomas, Mason and Tom, Dennis and Toomey, Emily and Tracy, Josh and Troyer, Matthias and Turley, Michelle and Turner, Matthew D. and Upadhyay, Shivendra and Urban, Ivan and Vaschillo, Alexander and Viazmitinov, Dmitrii and Vogel, Dominik and Wang, Zhenghan and Watson, John and Webster, Alex and Weston, Joseph and Williamson, Timothy and Winkler, Georg W. and van Woerkom, David J. and Wütz, Brian Paquelet and Yang, Chung Kai and Yu, Richard and Yucelen, Emrah and Zamorano, Jesús Herranz and Zeisel, Roland and Zheng, Guoji and Zilke, Justin and Zimmerman, Andrew},
	year = {2025},
}

@misc{AutoCASEOS,
  note={The technical details of the AutoCAS-EOS method will be the subject of a forthcoming publication.}
}

@misc{SparseIsometry,
  note={The technical details of the sparse isometry method for state preparation in \qdkchem{} will be the subject of a forthcoming publication.}
}

\appendix

\section{Example workflow code}
\label{appendix:example-code}

This appendix provides a complete example demonstrating a typical \qdkchem{} workflow, from molecular geometry specification through quantum phase estimation.
The example illustrates the modular design principles discussed in Section~\ref{sec:design}, showing how different workflow stages compose through uniform interfaces.

\begin{codeblock}[label={lst:workflow-example}]{Listing~\thetcbcounter: Complete \qdkchem{} workflow example.}
# QDK/Chemistry end-to-end example
# Ground state energy estimation via Quantum Phase Estimation

import numpy as np
from qdk_chemistry import algorithms as algo
from qdk_chemistry.data import SciWavefunctionContainer, Structure, Wavefunction
from qdk_chemistry.utils import compute_valence_space_parameters

# Import plugins for backends
import qdk_chemistry.plugins.qiskit
import qdk_chemistry.plugins.pyscf

# ------------------------------------------------------------
# Step 1: Define molecular structure
# ------------------------------------------------------------

# Structure: common representation for all algorithms
# Create stretched H2 molecule
structure = Structure(
    np.array([[0, 0, 0], [0, 0, 2.5]]),
    symbols=["H", "H"]
)

# ------------------------------------------------------------
# Step 2: Self-Consistent Field (SCF) calculation
# ------------------------------------------------------------

# Factory pattern: algo.create(category, [method])
# Common interface: run(input, **params) -> output
# Using the PySCF SCF backend; alternatives include "qdk"
scf_solver = algo.create("scf_solver", "pyscf")
scf_energy, scf_wfn = scf_solver.run(
    structure,
    charge=0,
    spin_multiplicity=1,
    basis_or_guess="sto-3g"
)

# ------------------------------------------------------------
# Step 3: Valence space selection
# ------------------------------------------------------------

num_val_e, num_val_o = compute_valence_space_parameters(scf_wfn, charge=0)
valence_selector = algo.create(
    "active_space_selector", "qdk_valence",
    num_active_electrons=num_val_e,
    num_active_orbitals=num_val_o
)
valence_wfn = valence_selector.run(scf_wfn)

# ------------------------------------------------------------
# Step 4: Hamiltonian construction
# ------------------------------------------------------------

ham_constructor = algo.create("hamiltonian_constructor")
active_hamiltonian = ham_constructor.run(valence_wfn.get_orbitals())

# ------------------------------------------------------------
# Step 5: Qubit Hamiltonian (Jordan-Wigner encoding)
# ------------------------------------------------------------

# Transform fermionic Hamiltonian to Pauli operators
# This example uses the Qiskit qubit mapper with Jordan-Wigner encoding
qubit_mapper = algo.create(
    "qubit_mapper", "qiskit", encoding="jordan-wigner"
)
qubit_hamiltonian = qubit_mapper.run(active_hamiltonian)

# ------------------------------------------------------------
# Step 6: Iterative Quantum Phase Estimation
# ------------------------------------------------------------

# Run CASCI to get multi-configuration wavefunction
n_alpha, n_beta = valence_wfn.get_active_num_electrons()
casci = algo.create("multi_configuration_calculator", "macis_cas")
casci_energy, casci_wfn = casci.run(active_hamiltonian, n_alpha, n_beta)

# Prepare trial state from top 2 determinants (truncated and renormalized)
trial_wfn = casci_wfn.truncate(max_determinants=2)

# Generate sparse state prep circuit (GF2 + X method)
state_prep = algo.create("state_prep", "sparse_isometry_gf2x")
state_prep_circuit = state_prep.run(trial_wfn)

# Create and configure IQPE
iqpe = algo.create(
    "phase_estimation",
    "iterative",
    num_bits=8,         # Number of phase bits
    evolution_time=0.5  # Evolution time
)

# Build the time evolution unitary via Trotter decomposition
evolution_builder = algo.create("time_evolution_builder", "trotter")

# Map controlled time evolution into circuit via Pauli sequence mapper
circuit_mapper = algo.create(
    "controlled_evolution_circuit_mapper",
    "pauli_sequence"
)

# Use QDK full state simulator to execute circuit and collect bitstrings
circuit_executor = algo.create("circuit_executor", "qdk_full_state_simulator")

# Run IQPE with the configured inputs
result = iqpe.run(
    state_preparation=state_prep_circuit,
    qubit_hamiltonian=qubit_hamiltonian,
    circuit_executor=circuit_executor,
    evolution_builder=evolution_builder,
    circuit_mapper=circuit_mapper,
)

# Extract energy from phase estimation result
qpe_energy = result.raw_energy

print("QPE Results (10-bit precision):")
print(f"Reference CASCI energy: {casci_energy:.6f} Ha")
print(f"QPE total energy: {qpe_energy:.6f} Ha")
print(f"Diff from CASCI: {abs(qpe_energy - casci_energy):.3e} Ha")
\end{codeblock}

\section{Acronyms}

\begin{acronym}
    \setlength{\itemsep}{0pt}
    \setlength{\parskip}{0pt}
    \setlength{\parsep}{0pt}
    \acro{ABI}{application binary interface}
    \acro{API}{application programming interface}
    \acro{AO}{atomic orbital}
    \acro{ASCI}{adaptive sampling configuration interaction}
    \acro{AVAS}{atomic valence active space}
    \acro{BFGS}{Broyden-Fletcher-Goldfarb-Shanno quasi-Newton optimization algorithm}
    \acro{BSE}{basis set exchange}
    \acro{CAS}{complete active space}
    \acro{CASCI}{complete active space configuration interaction}
    \acro{CASSCF}{complete active space self-consistent field}
    \acro{CCSD}{coupled cluster with single and double excitations}
    \acro{CCSD(T)}{coupled cluster with single and double excitations and perturbative triple excitations}
    \acro{CI}{configuration interaction}
    \acro{CPU}{central processing unit}
    \acro{DFT}{density functional theory}
    \acro{DIIS}{direct inversion in the iterative subspace}
    \acro{ECP}{effective core potential}
    \acro{ERI}{electron repulsion integral}
    \acro{FCI}{full configuration interaction}
    \acro{GDM}{geometric direct minimization}
    \acro{GGA}{generalized gradient approximation}
    \acro{GPU}{graphics processing unit}
    \acro{HF}{Hartree-Fock}
    \acro{HOMO}{highest occupied molecular orbital}
    \acro{HPC}{high-performance computing}
    \acro{JSON}{javascript object notation}
    \acro{LDA}{local density approximation}
    \acro{LUMO}{lowest unoccupied molecular orbital}
    \acro{MACIS}{Many-body Adaptive Configuration Interaction Solver}
    \acro{MC}{multi-configuration}
    \acro{MCSCF}{multi-configuration self-consistent field}
    \acro{MO}{molecular orbital}
    \acro{MP2}{second-order Møller-Plesset perturbation theory}
    \acro{QDK}{Quantum Development Kit}
    \acro{QPE}{quantum phase estimation}
    \acro{QPU}{quantum processing unit}
    \acro{RDM}{reduced density matrix}
    \acro{RHF}{restricted Hartree-Fock}
    \acro{RKS}{restricted Kohn-Sham}
    \acro{ROHF}{restricted open-shell Hartree-Fock}
    \acro{ROKS}{restricted open-shell Kohn-Sham}
    \acro{SCF}{self-consistent field}
    \acro{SCI}{selected configuration interaction}
    \acro{STO}{Slater-type orbital}
    \acro{UHF}{unrestricted Hartree-Fock}
    \acro{UKS}{unrestricted Kohn-Sham}
    \acro{VQE}{variational quantum eigensolver}
    \acro{VVHV}{valence virtual hard virtual}
\end{acronym}

\end{document}